\documentclass[journal]{IEEEtran}
\ifCLASSINFOpdf

\else

\fi

\usepackage{mathrsfs}
\usepackage{amsmath}
\usepackage{amsfonts}
\usepackage{amsthm}
\usepackage{amssymb}
\usepackage{graphicx}
\usepackage{subfigure}
\usepackage{indentfirst}
\usepackage{array}
\usepackage{epstopdf}
\usepackage{cite}
\usepackage{mathrsfs}
\usepackage{enumerate}
\usepackage{bm}
\usepackage[linesnumbered,ruled,vlined]{algorithm2e}
\usepackage{setspace}
\usepackage{multirow}
\usepackage{verbatim}
\usepackage{stfloats}
\usepackage{color}
\usepackage[table]{xcolor}
\usepackage{makecell}
\usepackage{cancel}
\usepackage{hyperref}

\hypersetup{colorlinks=true, linkcolor=blue, citecolor=blue}
\definecolor{lightgray}{rgb}{0.83, 0.83, 0.83}
\definecolor{revised}{HTML}{b71a3b}

\SetKwComment{Comment}{//}{}
\SetKwBlock{Begin}{Function}{end}

\begin{document}
	
	\title{SwinJSCC: Taming Swin Transformer for Deep Joint Source-Channel Coding}
	
	\author{Ke~Yang,~\IEEEmembership{Graduate Student Member,~IEEE},
		Sixian~Wang,~\IEEEmembership{Member,~IEEE},
		Jincheng~Dai,~\IEEEmembership{Member,~IEEE},
		Xiaoqi~Qin,~\IEEEmembership{Senior Member,~IEEE},
		Kai~Niu,~\IEEEmembership{Member,~IEEE},
		and Ping~Zhang,~\IEEEmembership{Fellow,~IEEE}
		
		\thanks{This paper has been partially presented in IEEE ICASSP 2023 \cite{yang2023witt}.}

		\thanks{This work was supported in part by the National Natural Science Foundationof China under Grant 62293481, Grant 62371063, and Grant 92067202, inpart by the Beijing Natural Science Foundation under Grant L232047, Grant 4222012, in part by Program for Youth Innovative Research Team of BUPT under Grant 2023QNTD02. (\emph{Corresponding author: Jincheng Dai.})}
		
		\thanks{Ke Yang, Sixian Wang, Jincheng Dai, and Kai Niu are with the Key Laboratory of Universal Wireless Communications, Ministry of Education, Beijing University of Posts and Telecommunications, Beijing 100876, China (e-mail: daijincheng@bupt.edu.cn).}
		
		\thanks{Xiaoqi Qin and Ping Zhang are with the State Key Laboratory of Networking and Switching Technology, Beijing University of Posts and Telecommunications, Beijing 100876, China.}
		
		\thanks{Our project and open source code are available at: \href{https://github.com/semcomm/SwinJSCC}{https://github.com/semcomm/SwinJSCC}.}
	}
	
	\maketitle

	\begin{abstract}
		
		As one of the key techniques to realize semantic communications, end-to-end optimized neural joint source-channel coding (JSCC) has made great progress over the past few years. A general trend in many recent works pushing the model adaptability or the application diversity of neural JSCC is based on the convolutional neural network (CNN) backbone, whose model capacity is yet limited, inherently leading to inferior system coding gain against traditional coded transmission systems. In this paper, we establish a new neural JSCC backbone that can also adapt flexibly to diverse channel conditions and transmission rates within a single model, our open-source project aims to promote the research in this field. Specifically, we show that with elaborate design, neural JSCC codec built on the emerging Swin Transformer backbone achieves superior performance than conventional neural JSCC codecs built upon CNN, while also requiring lower end-to-end processing latency. Paired with two spatial modulation modules that scale latent representations based on the channel state information and target transmission rate, our baseline SwinJSCC can further upgrade to a versatile version, which increases its capability to adapt to diverse channel conditions and rate configurations. Extensive experimental results show that our SwinJSCC achieves better or comparable performance versus the state-of-the-art engineered BPG + 5G LDPC coded transmission system with much faster end-to-end coding speed, especially for high-resolution images, in which case traditional CNN-based JSCC yet falls behind due to its limited model capacity.
		
	\end{abstract}
	
	
	\IEEEpeerreviewmaketitle
	
	\section{Introduction}\label{section_introduction}

		\subsection{Background and Related Work}

		Guided by the Shannon separation principle \cite{shannon1948mathematical}, traditional communication systems have been designed using the separation approach, which optimize source and channel modules independent of each other. The separation approach optimal is theoretically in the asymptotic limit of infinitely long source and channel blocks and unlimited delay. However, the assumptions on which separation theory is based may not hold in a practical system, which leads to the development of joint source-channel coding (JSCC). JSCC can greatly improve the system performance when there are, for example, stringent end-to-end delay constraints or implementation concerns \cite{fresia2010joint}. 
	
		Recently, end-to-end optimized neural or deep learning-based JSCC (deep JSCC) for data transmission has emerged as an active research area in semantic communications \cite{gunduz2022beyond, dai2022communication, bourtsoulatze2019deep, xu2021wireless, kurka2021bandwidth, yang2022deep, zhang2023predictive, yuan2023channel, bian2023deepjscc}. Specifically, for image transmission tasks, current deep JSCC\cite{bourtsoulatze2019deep} and its variants\cite{xu2021wireless, kurka2021bandwidth, yang2022deep, zhang2023predictive} using convolutional neural networks (CNN) backbone can yield end-to-end image transmission performance surpassing classical separation-based methods (JPEG/JPEG2000/BPG combined with advanced channel codes)\cite{bourtsoulatze2019deep}. 
	
		Bourtsoulatze et al. \cite{bourtsoulatze2019deep} have proposed the first CNN-based deep JSCC scheme outperforms separation-based digital transmission scheme at low signal-to-noise ratio (SNR) and channel bandwidth regimes, especially for sources of small dimensions, e.g., tiny CIFAR10 image ($32\times 32$ resolution) dataset \cite{CIFAR10}, highlighted the efficacy of their approach. Later, Xu et al. \cite{xu2021wireless} proposed  Attention DL-based JSCC that constructs a single network capable of handling a range of SNR values for better model adaptability. Yuan et al. \cite{yuan2023channel} further improved the method, achieving exceptional performance across various SNR levels during transmission without relying on channel prior information. Yang et al. \cite{yang2022deep} achieved adaptive rates according to different channel SNR and image contents. However, these previous works employed CNN networks to focus primarily on low-resolution image datasets, employing variations in parameter configurations and network structures to improve performance. In addition, their adaption strategies involve either the channel state or the target rate, or none. In this context, Zhang et al. \cite{zhang2023predictive} attempted to develop a flexible approach adapting to both channel SNR and transmission rate simultaneously, but with a clear expense of transmission performance degradation.
		
		\subsection{Motivation and Contribution}
		
		With the escalation in image resolution, the aforementioned CNN-based deep JSCC models generally hard to learn the hierarchical features and image details, leading to clear performance degradation. This phenomenon can be partially attributed to the limited representation capability of CNN, thus one of the fundamental aspects to enhance the transmission performance of deep JSCC models revolves around improving the model capacity. Meanwhile, it is crucial to account for the effects of both varying channel states and various transmission rates. 
		
\begin{figure}[t]
			\setlength{\abovecaptionskip}{0.cm}
			\setlength{\belowcaptionskip}{-0.cm}
			\centering{\includegraphics[scale=0.4]{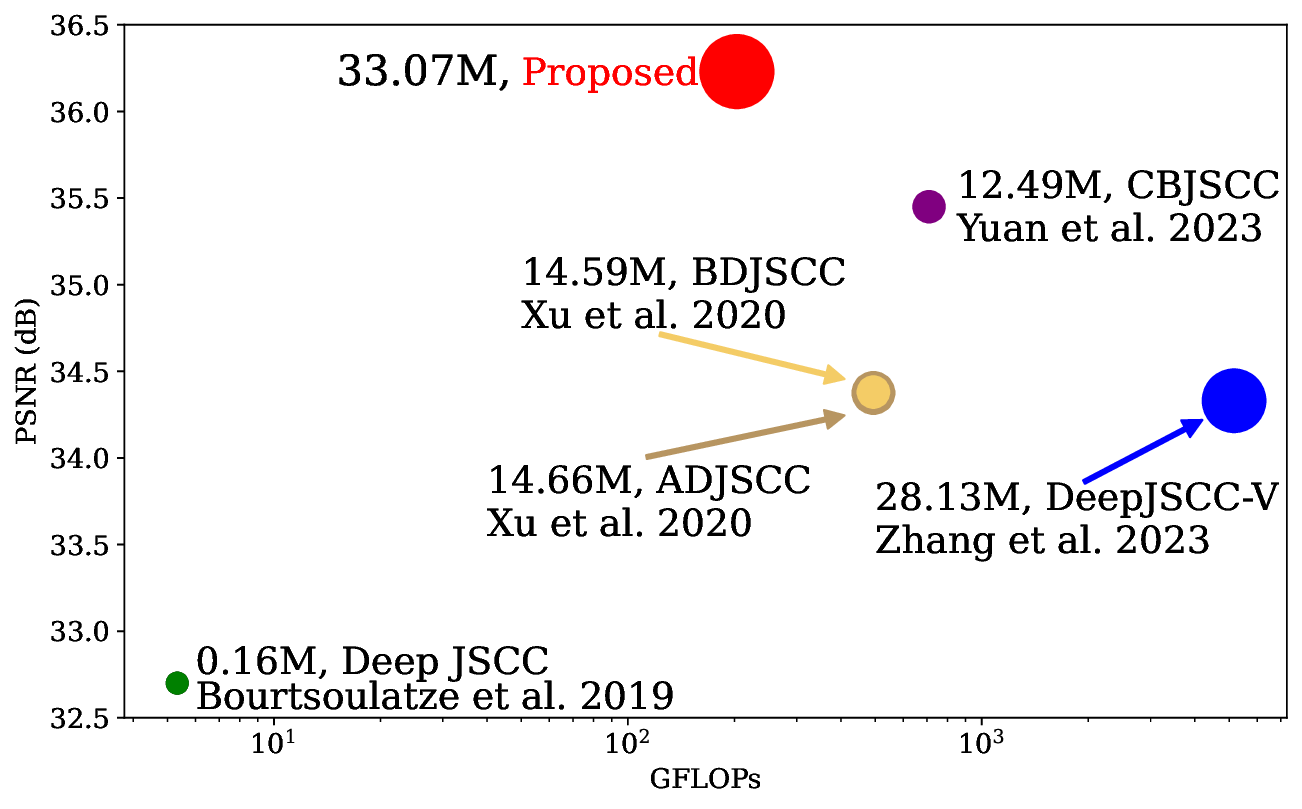}}
			\caption{Reconstructed image PSNR performance versus processing complexity of different methods for Kodak dataset over AWGN channel where SNR = 7dB and channel bandwidth ratio (CBR) = 1/6. Complexity measures include the Floating Point Operations and the size of model parameters. The increased model parameters are a result of incorporating two adaptation types within our proposed and the left-top is better.}\label{Fig1}
		\end{figure}

		Therefore, in this paper, we establish a new network backbone to upgrade deep JSCC, our open-source framework aims to promote research in this field. Our method is based on the emerging Transformer architecture that contains no built-in inductive prior to the locality of interactions and is free to learn complex contextual relationships among its inputs. The global attention mechanism inside Transformer enables a closer connection among image patches, which further contributes to the stronger capability to combat channel noise and interference. A brief performance demonstration, as illustrated in Fig. \ref{Fig1}, compares our method with other existing approaches in terms of reconstructed image PSNR and processing complexity. With the elaborate design, our model integrates the advantages of Transformer \cite{vaswani2017attention} into the deep JSCC framework, resulting in improved transmission performance while simultaneously holding acceptable latency and computational complexity.
		
		Specifically, this paper takes the lead to investigate a new architecture named SwinJSCC to address the limitations of CNN-based JSCC methods by integrating the Swin Transformer \cite{dosovitskiy2020image,liu2021swin} into the deep JSCC framework. The Swin Transformer, which constructs hierarchical feature maps in the latent space and has linear computational complexity to image size, is utilized as the key component of our proposed framework. Although the Swin Transformer has been widely investigated in vision analysis tasks, it has not been applied to JSCC, particularly lacking an elaborate design to handle the effects of varying channel states and various transmission rates. Thus, a naive alternation of CNN as Swin Transformer in JSCC cannot yet achieve considerable performance gain. We tackle this by developing a flexible and comprehensive model capable of simultaneously adapting to both channel SNR and transmission rate while maintaining the desired performance. 
		
		As we can see, JSCC for wireless image/video transmission applications is a typical rate allocation problem between source coding and channel coding \cite{fresia2010joint}, which has been studied for a long time. Recent deep JSCC works inherently \emph{learn to find} a near-optimal solution through stochastic optimization methods with different models. However, due to limited model capacity, under a variety of channel conditions and source data, finding all these near-optimal rate allocation solutions within a single model is a very challenging task. Our SwinJSCC with considerably increased model capacity can logistically provide a possibility for solving such optimization problems under diverse conditions. To this end, we propose two plug-in modules into SwinJSCC, i.e., Channel ModNet and Rate ModNet, to enable a single model to handle various transmission rates and channel conditions while also guaranteeing a stable transmission quality. Specifically, the Channel ModNet is responsible for making SwinJSCC aware of the channel condition, thus a reasonable rate allocation solution under each specific channel state can be implicitly found. The Rate  ModNet realizes very flexible transmission bandwidth control through a learnable mask on the latent representations. These two modules jointly facilitate a versatile SwinJSCC framework. We verify the superiority of SwinJSCC via extensive experiments. Overall, the new SwinJSCC backbone improves the efficiency of wireless image transmission and provides a comprehensive solution to handle various channel conditions.
		
		In this paper, the main contributions of this paper are listed as follows:
		\begin{itemize}
			\item [1)] \emph{Swin Transformer-Based JSCC Framework:} Combining the advantages of Transformer and JSCC, we propose a SwinJSCC scheme that utilizes the Swin Transformer as a new backbone for JSCC to improve the model capacity and transmission performance.
			\item [2)] \emph{SNR and Rate Adaptation:} To enhance the robustness of SwinJSCC, we propose two plug-in modules, Channel ModNet and Rate ModNet, which are optimized for various channel conditions in communication scenarios, allowing a single model to adapt to different channel states and transmission rates for flexible wireless transmission.
		\end{itemize}

		The remainder of this paper is organized as follows. In section \ref{section_system_model}, we first review the system model of deep JSCC. Then, in section \ref{section_proposed_method}, we introduce the overview of the SwinJSCC framework. Section \ref{adaptive mechanism} is dedicated to the details of the Channel ModNet and Rate ModNet. Section \ref{Experimental} provides the experiment results and a direct comparison of several methods to quantify the performance gain of the proposed method. Finally, section \ref{conclusion} concludes this paper.
		
		\emph{Notational Conventions:} Throughout this paper, lowercase letters (e.g., $x$) denote scalars, bold lowercase letters (e.g., $\boldsymbol{x}$) denote vectors. Bold uppercase letters (e.g., $\boldsymbol{X}$) denote matrices, and $\boldsymbol{I}_m$ denotes an $m$-dimensional identity matrix. $\log(\cdot)$ denotes the logarithm to base 2. $p_x$ denotes a probability density function (pdf) with respect to the continuous-valued random variable $x$, and $P_{\bar x}$ denotes a probability mass function (pmf) for the discrete-valued random variable $\bar x$. In addition, $\mathbb{E} [\cdot]$ denotes the statistical expectation operation, and $\mathbb{R}$ denotes the real number set. Finally, $\mathcal{N}(x|\mu, \sigma^2) \triangleq (2\pi \sigma^2)^{-1/2} \exp(-(x - \mu)^2/(2\sigma^2))$ denotes a Gaussian function.

	\section{System Model}\label{section_system_model}

	Consider the following lossy end-to-end transmission scenario. Alice is drawing an image from the source, denoting as an $m$-dimensional vector $\boldsymbol{x}$, whose probability is given as $p_{\boldsymbol{x}}( \boldsymbol{x} )$. Alice concerns how to map $\boldsymbol{x}$ to a $k$-dimensional vector $\boldsymbol{y}$, where $k$ is referred to as the \emph{channel bandwidth cost}, and $R = k/m$ is referred to as the \emph{channel bandwidth ratio (CBR)} that is typically lower than 1. Then, Alice transmits $\boldsymbol{y}$ to Bob via a realistic communication channel, who uses the received information $\boldsymbol{\hat y}$ to reconstruct an approximation to $\boldsymbol{x}$.
	
\begin{figure}[t]
		\setlength{\abovecaptionskip}{0.cm}
		\setlength{\belowcaptionskip}{-0.cm}
		\centering{\includegraphics[scale=0.36]{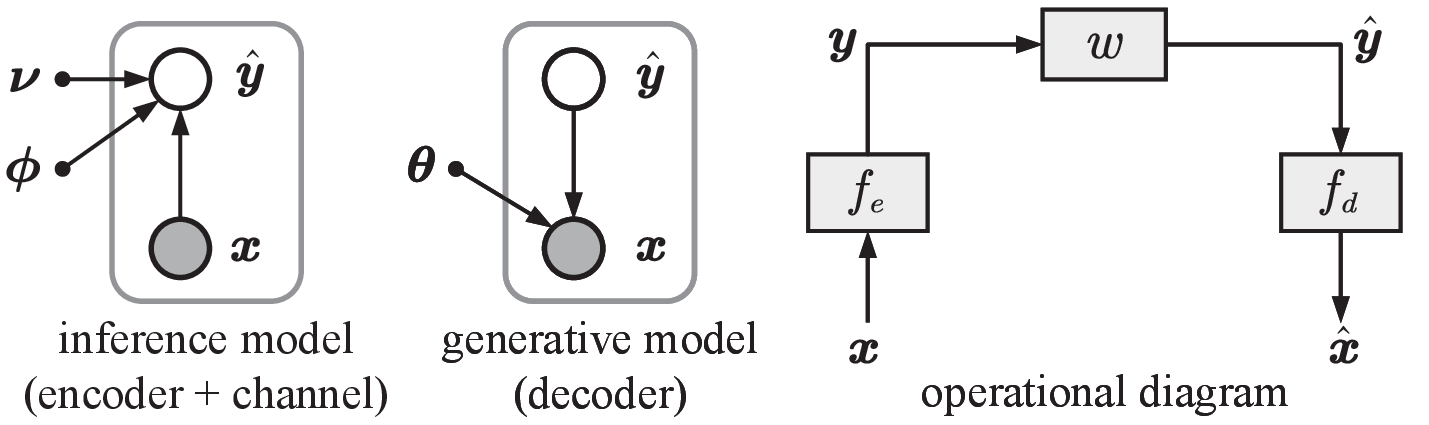}}
		\caption{Left: representation of a deep JSCC encoder combined with the communication channel as an inference model and corresponding decoder as a generative model. Nodes denote random variables or parameters, and arrows show conditional dependence between them. Right: diagram showing the operational structure of the deep JSCC transmission model. Arrows indicate the data flow, and boxes represent the coding functions of data and channels.}\label{Fig2}
	\end{figure}

	Different from traditional separation-based source and channel coding methods \cite{BPG,witten1987arithmetic, LDPC_5G, Polar}, in deep JSCC \cite{bourtsoulatze2019deep}, the source vector $\boldsymbol x \in {\mathbb R}^m$, is mapped to a vector of continuous-valued channel input symbols $\boldsymbol y \in {\mathbb R}^k$ via an encoding function ${\boldsymbol y} = f_{e}( {\boldsymbol x}; {\bm \phi} )$, where the encoder was usually parameterized as a convolutional neural network (CNN) with parameters ${\bm \phi}$. Then, the analog sequence $\boldsymbol{y}$ is directly sent over the wireless channel. The channel introduces random corruptions to the transmitted symbols, denoted as a function $W( \cdot; \bm{\nu} )$, and the channel parameters are encapsulated in $\bm{\nu}$. Accordingly, the received sequence is ${\boldsymbol{\hat y}} = W( \boldsymbol{y} ; \bm{\nu} )$, whose transition probability is ${{p_{{\boldsymbol{\hat y}}| {\boldsymbol{y}} }}( {{\boldsymbol{\hat y}}| \boldsymbol{y} } )}$. In this paper, we consider the most widely used AWGN channel model such that the transfer function is ${\boldsymbol{\hat y}} = W( \boldsymbol{y}; \sigma_n ) = \boldsymbol{y} + \boldsymbol{n}$ where each component of the noise vector $\boldsymbol{n}$ is independently sampled from a Gaussian distribution, i.e., $\boldsymbol{n} \sim \mathcal{N}(0, {\sigma_n^2}{\boldsymbol{I}}_k)$, where ${\sigma_n^2}$ is the average noise power. Changing the channel transition function can also similarly incorporate other channel models. The receiver also comprises a parametric function ${{\boldsymbol {\hat x}}} = f_{d}( {{\boldsymbol {\hat y}}}; {\bm \theta} )$ to recover the corrupted signal ${\boldsymbol{\hat y}}$ as ${{\boldsymbol {\hat x}}}$, where $f_d$ can also be a format of CNN \cite{bourtsoulatze2019deep}. As analyzed in \cite{saidutta2021joint}, the deep JSCC can also be modeled as a variational autoencoder (VAE) \cite{kingma2013auto}. As shown in the left panel of Fig. \ref{Fig2}, the noisy sequence $\boldsymbol{\hat y}$ can be viewed as a sample of latent variables in the generative model. The deep JSCC decoder acts as the generative model (``generating'' the reconstructed source from the latent representation) that transforms a latent variable with some predicted latent distribution into an unknown data distribution. The deep JSCC encoder combined with the channel is linked to the inference model (``inferring'' the latent representation from the source data). The whole operation is shown in the right panel of Fig. \ref{Fig2}. The encoder and decoder functions are jointly learned to minimize the average
	\begin{equation}\label{eq_1}
		\left( {{{\bm{\phi}}^*},{{\bm{\theta}}^*}} \right) = \arg \mathop {\min }\limits_{{\bm{\phi}} ,{\bm{\theta}} } {{\mathbb{E}}_{\boldsymbol{x}\sim{p_{\boldsymbol{x}}}}}{{\mathbb{E}}_{{{\boldsymbol{\hat{y}}} \sim p_{{\boldsymbol{\hat{y}}} | {\boldsymbol{x}} }}}}\left[ {d\left( {{\boldsymbol{x}},{\boldsymbol{\hat{x}}}} \right)} \right],
	\end{equation}
	where $d(\cdot)$ denotes the distortion loss function.
	
	In this paper, in terms of the image semantic transmission task, the distortion function $d (\cdot)$ for image quality assessment (IQA) between $\boldsymbol{x}$ and $\boldsymbol{\hat x}$ will be chosen as both the objective metric and the perceptual metrics aligned with human quality ratings. As for the objective metric, the codec parameters of deep JSCC methods are usually adjusted to minimize the MSE, the simplest of all fidelity metrics, even though it has been widely criticized for its poor correlation with human perception of image quality \cite{girod1993s}. In this case, the distortion function is $d\left( {{\boldsymbol{x}},{\boldsymbol{\hat{x}}}} \right) = \| \boldsymbol{x} - \boldsymbol{\hat x} \|_2^2$, and the IQA metric is the peaks-signal-to-noise ratio (PSNR) \cite{ding2021comparison}. The PSNR metric has been widely used in image processing applications and provides a simple yet effective measurement of image quality.
	
	In image restoration, prior efforts toward perceptual optimization using the structural similarity (SSIM) index instead of mean squared error (MSE) have shown perceptual improvements. The multi-scale SSIM (MS-SSIM) \cite{wang2003multiscale} provides greater versatility than single-scale SSIM, making it suitable for a broader viewing range. This method involves decomposing images into Gaussian pyramids, computing contrast and structure similarities at each scale, and luminance similarity at the coarsest scale. In this paper, we adopt MS-SSIM, a well-established perceptual metric, as the distortion function $d(\cdot)$ during model training and as the IQA metric for model testing.

	\section{The Proposed SwinJSCC Framework}\label{section_proposed_method}
	
	In this section, we present the SwinJSCC framework for wireless image transmission. Our exposition proceeds in three main parts. First, we conduct an in-depth analysis of the impact of model capacity on transmission performance. Second, we illustrate the capacity-enhanced architecture of SwinJSCC, which concerns channel SNR and rate adaptation image transmission. Finally, we further propose two new single adaptive SwinJSCC schemes to simplify the training process.

	\subsection{Analysis on JSCC Model Capacity and Representation}
	
	As analyzed in \cite{dosovitskiy2020image,liu2021swin}, $\boldsymbol{y}$ is indeed the learned latent representation of the source image $\boldsymbol{x}$. While in deep JSCC model, with respect to traditional error correction coding, the latent representation $\boldsymbol{y}$ can also combat the channel noise like the channel coded sequence. In the computer vision (CV)-related image transmission task, the design of both encoder $f_e$ and decoder $f_d$ is critical in determining the semantic level of the learned latent representations. Typically, a CNN network consists of $s$ stages, where each stage's convolutional layers share the same structure. Let $\boldsymbol{x}_i$ denotes the feature with a spatial size of $(H_i, W_i)$ and $C_i$ channels in the $i$ stage, the $i$ stage comprises a stack of $L_i$ identical convolutional layers $F_i$ \cite{tan2019efficientnet}. Thus, the entire convolutional network $\mathcal{N}$ can be represented as follows:
	\begin{equation}\label{eq_2}
		\mathcal{N} = \mathop {\odot}\limits_{i=1,...,s} {F_i}^{L_i}(\boldsymbol{x}_{<H_i,W_i,C_i>})
	\end{equation}
	where ${F_i}^{L_i}$ denotes layer $F_i$ is repeated $L_i$ times in stage $i$, $<H_i, W_i, C_i>$ denotes the shape of input tensor $\boldsymbol{x}$ of stage $i$.
	
	\begin{figure}[t]
		\setlength{\abovecaptionskip}{0.cm}
		\setlength{\belowcaptionskip}{-0.cm}
		\centering{\includegraphics[scale=0.6]{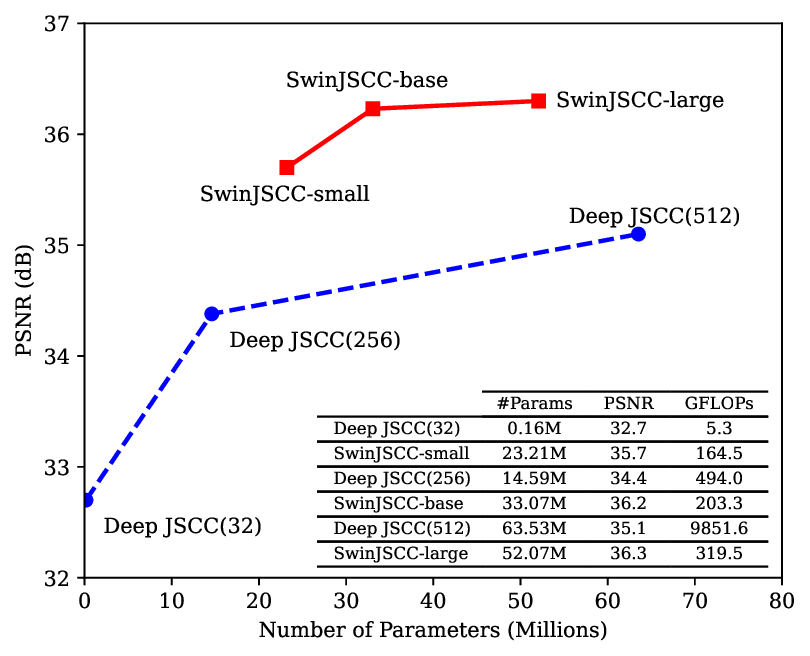}}
		\caption{Model Size vs. Performance. All numbers are for the Kodak dataset over AWGN channel where SNR = 7dB and channel bandwidth ratio (CBR) = 1/6. Deep JSCC($C$) denotes a CNN architecture comprising multiple convolutional layers, each employing $C$ channels. SwinJSCC-small, SwinJSCC-base, and SwinJSCC-large are three different model size versions of our proposed models.}\label{Fig3}  
	\end{figure}
	
	Therefore, in a CNN network, the capacity of the model is closely related to the model width, model depth, and resolution of the input image \cite{tan2019efficientnet}. Among these factors, model width (number of channels per stage) is the most significant parameter affecting model capacity. Increasing the number of channels per layer can indeed increase the number of parameters and feature dimensions in the network, thereby enhancing the model's capacity. More channels can enable the model to learn richer feature representations, leading to improved performance on complex tasks. Fig. \ref{Fig3} illustrates the impact of increasing model width on the variation in model performance. The result shows that increasing the width of a CNN-based model does indeed lead to an increase in model complexity and computational demands, and there also exists a saturation point in terms of performance gains, where further increases in model width may not result in substantial improvements. Therefore, we seek a new backbone approach to realizing deep JSCC for wireless image transmission.
	
	To address the limited capacity of deep JSCC methods for wireless image transmission, several approaches\cite{hu2019local,ramachandran2019stand,zhao2020exploring} have been explored to enhance the network capacity. One such approach is to replace some or all of the spatial convolution layers in the CNN with self-attention layers, which have been successful in NLP. However, these approaches require higher memory access costs resulting in significant latency compared to convolutional networks. Another method is to augment a standard CNN architecture with self-attention layers or Transformers, which can encode distant dependencies or heterogeneous interactions to complement backbones\cite{cao2019gcnet} or head networks\cite{hu2018relation}. Recently, the encoder-decoder design in Transformer has been applied for many CV tasks \cite{carion2020end,chi2020relationnet++}. Inspired by them, we aim to tame Transformers for deep JSCC to realize much higher efficient image semantic transmission.

	\begin{figure}[t]
		\setlength{\abovecaptionskip}{0.cm}
		\setlength{\belowcaptionskip}{-0.cm}
		\centering{\includegraphics[scale=0.36]{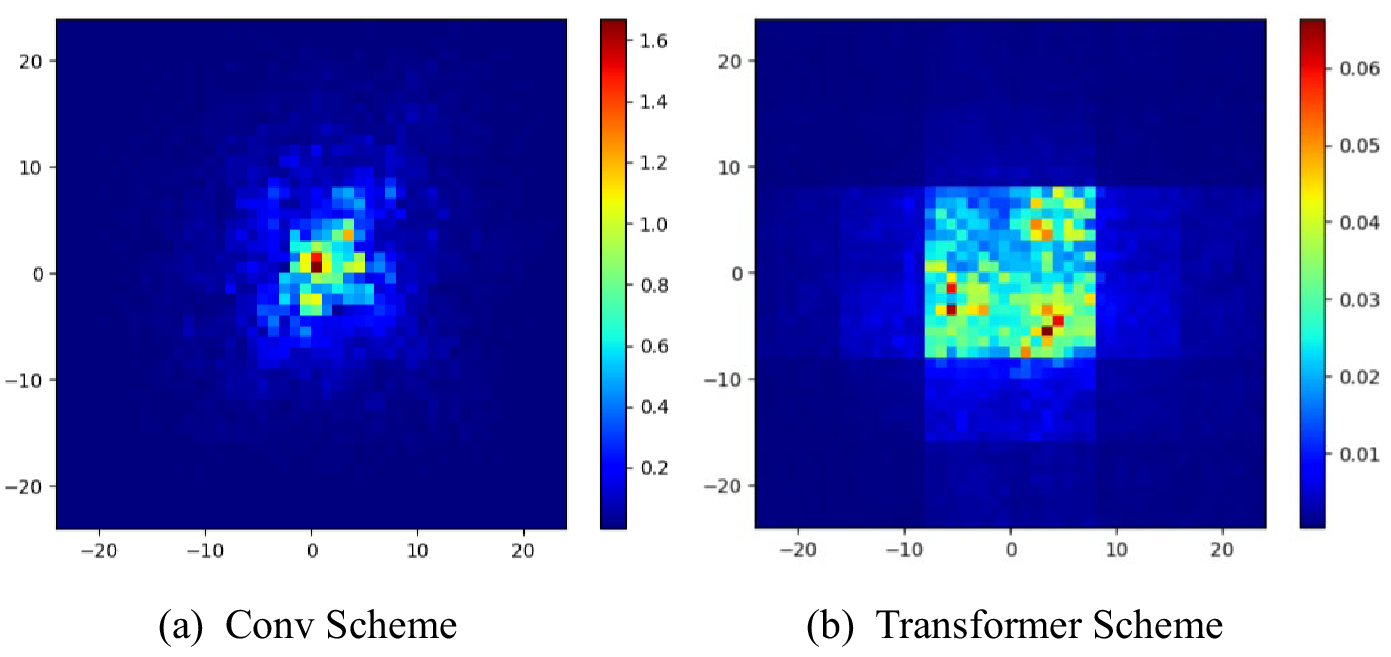}}
		\caption{Comparison of the effective receptive field (ERF) between the encoders $f_e$ of Conv scheme and Transformer scheme. ERF is visualized as absolution gradients of the center pixel in the latent (i.e., $d\boldsymbol{y}/d\boldsymbol{x}$) with respect to the input image, specifically 24 Kodak images cropped to $512 \times 512$ for image codecs. The plot shows the close-up of the gradient maps averaged over all channels in each input of test images.}\label{Fig4}  
	\end{figure}
	
	Fig. \ref{Fig4} shows the effective receptive field (ERF) of the Transformer encoder and Convs encoder. The ERF represents the perception range of each neuron in the neural network towards the input data, determining the scope of local and global information that the network can learn. A larger receptive field enables the model to capture global context information and the long-term dependencies of source input, resulting in improved model performance. Transformer-based models typically exhibit a wider ERF, which facilitates capturing long-term correlations and yields superior performance in image transmission tasks. Moreover, Transformer-based models also require meticulous consideration of the design of model width and depth. As depicted in Fig. \ref{Fig3}, we are employing three distinct configurations for training and testing in order to derive an optimal design solution.
	
	Our work on deep JSCC using Transformer-based architectures is most closely related to the Vision Transformer (ViT) \cite{dosovitskiy2020image} and its derivatives\cite{han2021Transformer,wang2021pyramid}. The architecture of ViT is still far from satisfying the requirements of dense vision tasks or when the input image resolution is high due to its low-resolution feature maps and the quadratic increase in complexity with image size. Accordingly, a direct application of ViT in deep JSCC will also result in unsatisfied performance on high CBR $R$ or high-resolution source images. In this paper, guided by the emerging Swin Transformer architecture \cite{liu2021swin}, an improved ViT variant, we propose new deep JSCC methods that achieve an excellent speed-performance trade-off versus existing methods. Our efficient approach achieves excellent wireless image transmission performance on objective or perceptual metrics, e.g., SNR, MS-SSIM, etc.
	
	\subsection{The Overall Architecture of SwinJSCC}
	
	\begin{figure*}[t]
		\setlength{\abovecaptionskip}{0.cm}
		\setlength{\belowcaptionskip}{-0.cm}
		\centering{\includegraphics[scale=0.4]{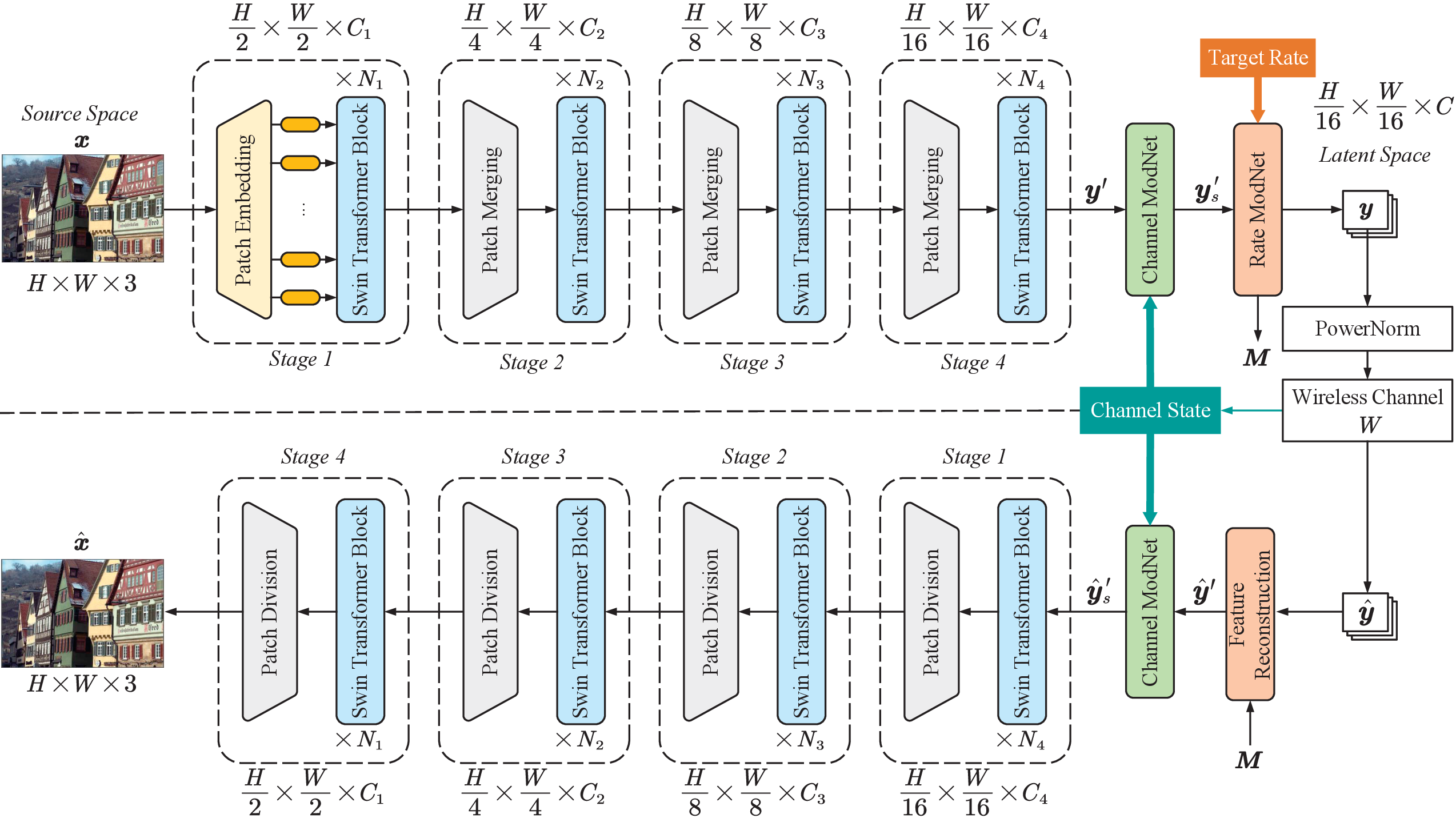}}
		\caption{The overall architecture of our SwinJSCC for wireless image transmission.}\label{Fig5}
	\end{figure*}
	
	\begin{figure*}[t]
		\setlength{\abovecaptionskip}{0.cm}
		\setlength{\belowcaptionskip}{-0.cm}
		\centering{\includegraphics[scale=0.4]{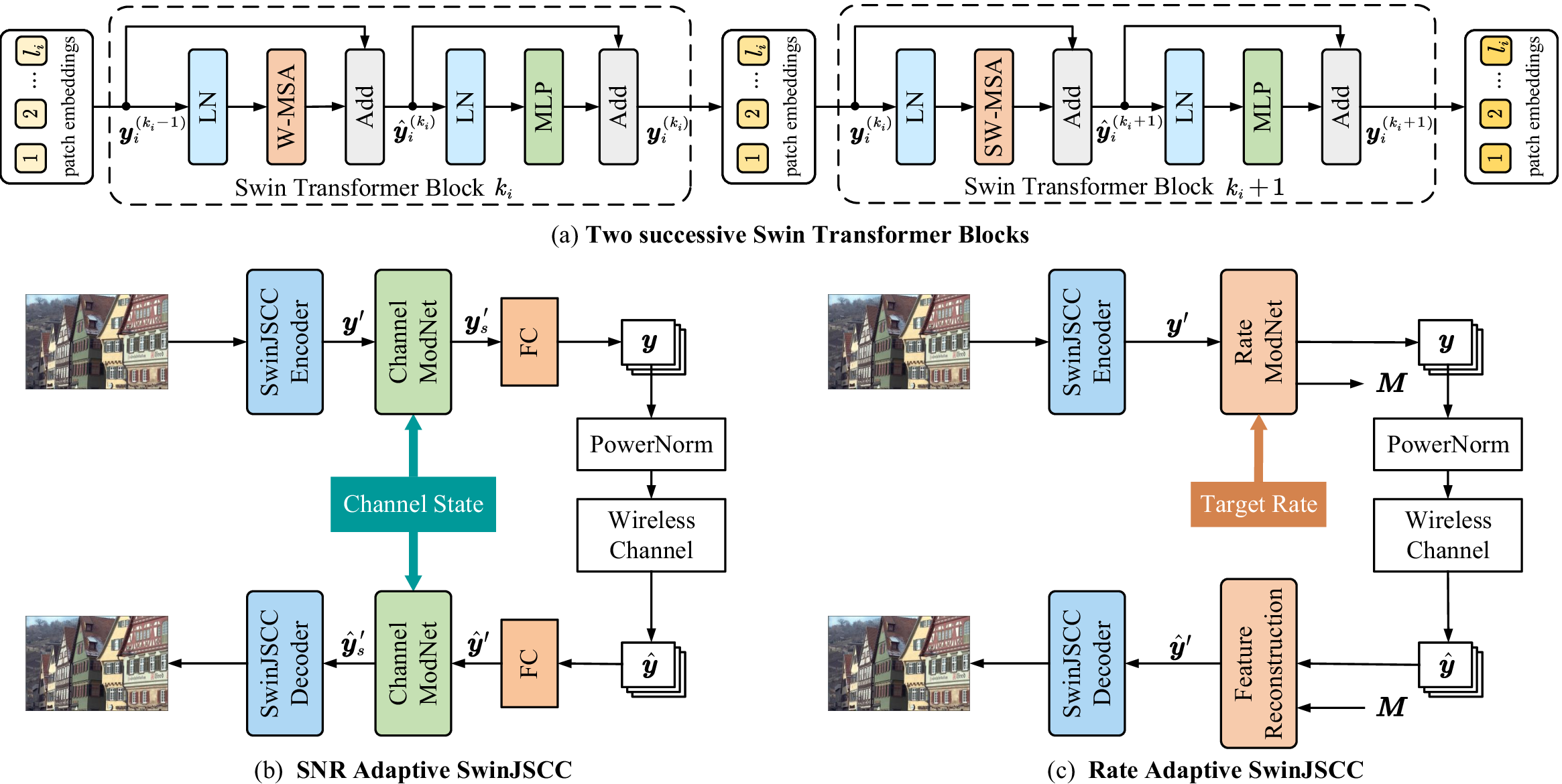}}
		\caption{(a) Two successive Swin Transformer Blocks. (b) The overall architecture of the proposed SNR Adaptive SwinJSCC scheme for wireless image transmission. (c) The overall architecture of the proposed Rate Adaptive SwinJSCC scheme for wireless image transmission.}\label{Fig6}
	\end{figure*}

	An overview of the proposed SwinJSCC architecture for wireless image transmission is presented in Fig. \ref{Fig5}. The input RGB image source $\boldsymbol{x} \in \mathbb{R}^{H\times W\times3}$ is partitioned into $l_1=\frac{H}{2} \times \frac{W}{2}$ non-overlapping patches, which are regarded as tokens and arranged in a sequence $( x_1, \dots, x_{l_1} )$ by following a left-to-right, top-to-bottom order. Subsequently, $N_1$ Swin Transformer blocks are applied to these $l_1$ tokens \cite{liu2021swin}. The number of patch embeddings output from these Swin Transformer blocks remains the same as $l_1=\frac{H}{2} \times \frac{W}{2}$. We collectively refer to these $N_1$  Swin Transformer blocks and the patch embedding layer as “stage 1”. As demonstrated in Fig. \ref{Fig6}(a), the Swin Transformer block operates on image patches. It incorporates the standard multi-head self-attention (MSA) module and feed-forward networks to process \cite{liu2021swin}. The shifted window-based self-attention mechanism allows the model to capture long-range dependencies within the image. It divides the image into a grid of windows and applies self-attention within each window.
	
	To construct a hierarchical representation, the number of tokens is gradually reduced via patch merging layers as the network delves deeper. Specifically, neighboring embeddings output from stage 1 are merged by a patch merging operation in stage 2, and the resulting concatenated embeddings of size $4C_1$ are reduced to size $C_2$. Subsequently, $l_2=\frac{H}{4} \times \frac{W}{4}$ patch embedding tokens with higher-resolution are fed into $N_2$ Swin Transformer blocks. As depicted in Fig. \ref{Fig5}, each stage consists of a down-sampling patch merging layer followed by several Swin Transformer blocks. The process above constitutes two stages in total. In this way, the proposed model remarkably improves model capacity as it captures long-range dependencies, exploits global information, and efficiently learns complex details in high-resolution images.
	
	The proposed encoder $f_e$, which consists of SwinJSCC encoder, Channel ModNet, and Rate ModNet, is designed to handle source images with high resolution and learn from the varying characteristics of the transmission channel. The number of stages in $f_e$ can be set according to the input image size. In this paper, we adopt a four stages $f_e$ encoder to generate a semantic latent representation $\boldsymbol{y}^\prime \in \mathbb{R}^{\frac{H}{16} \times \frac{W}{16} \times C_4}$ as shown in Fig. \ref{Fig5}. The latent representation captures the semantic features of the source image, allowing it to combat the effects of channel noise and fading, etc. After multiple processing stages,  we integrate two ModNet modules into the encoder, enabling enhanced adaptability to the varying characteristics of transmission channels. All patch embeddings are fed into the Channel ModNet module to adapt to changing channel state. Subsequently, a Rate ModNet is employed to adjust the embedding size to match the CBR $R$, defined as $R = C/( 2 \times 3 \times 2^i \times 2^i )$, where $i$ denotes the number of stages. The Rate ModNet module generates a binary vector mask $\boldsymbol{M}$ with the same resolution as $\boldsymbol{y}_s^\prime$. The final feature map $\boldsymbol{y}$ represents the semantic latent representation of the input image $\boldsymbol{x}$.
	
	Before transmitting $\boldsymbol{y}$ into the wireless channel, a power normalization operation described in \cite{bourtsoulatze2019deep} is carried out on the produced feature map $\boldsymbol{y}$. Subsequently, the resulting analog feature map is directly transmitted over the wireless channel. We consider the general fading channel model with transfer function $\boldsymbol{\hat{y}} = W( \boldsymbol{y}; \boldsymbol{h} ) = \boldsymbol{h} \odot \boldsymbol{y} + \boldsymbol{n}$, where $\odot$  represents the element-wise product, $\boldsymbol{h}$ denotes the channel state information (CSI) vector, and $\boldsymbol{n}$ means the noise vector, whose components are independently drawn from a Gaussian distribution, i.e., $\boldsymbol{n} \sim \mathcal{N}(0, {\sigma_n^2}{\boldsymbol{I}}_k)$, where ${\sigma_n^2}$ is the average noise power. The feature map $\boldsymbol{y}$ generated here constitutes the semantic latent representation of the source image $\boldsymbol{x}$.
	
	The proposed decoder $f_d$ follows a symmetric architecture with encoder $f_e$, which includes the feature reconstruction, Channel ModNet, the patch division operation for up-sampling, and Swin Transformer. The feature reconstruction module first pads the masked positions from the received symbol $\boldsymbol{\hat{y}}$ with zero values using the side information mask $\boldsymbol{M}$. Then, it reconstructs the original image from the noisy latent representation. It is noteworthy that, as shown in Fig. \ref{Fig5}, both the encoder and decoder modules incorporate the feedback channel signal-to-noise ratio (SNR) and the predetermined CBR $R$ as special tokens sent into the Channel ModNet and Rate ModNet to adapt to the varying channel SNRs and achieve the target transmission rates. The entire decoding process can be expressed as
	\begin{equation}\label{eq_3}
		\boldsymbol{\hat{x}} = f_{d}(\boldsymbol{\hat{y}}, \boldsymbol{M}) = f_{d}(W(f_{e}(\boldsymbol{x}, \text{SNR}, R);\boldsymbol{h}), \boldsymbol{M}),
	\end{equation}
	where $\boldsymbol{\hat{x}}$ is the reconstruction image.
	
	Based on the Channel ModNet and Rate ModNet modules, we have built a versatile SwinJSCC scheme. The proposed scheme offers a promising solution for end-to-end image transmission, and it is capable of enhancing the represented capacity of the model and supports both SNR adaptive and rate adaptive strategies for wireless image transmission. During the whole model training, we minimize the following loss to encourage improving the image reconstruction quality, and the training loss function of the whole system is
	\begin{equation}\label{eq_4}
		\mathop{\min }\limits_{{\bm{\phi}} ,{\bm{\theta}} } {{\mathbb{E}}_{\boldsymbol{x}\sim{p_{\boldsymbol{x}}}}}{{\mathbb{E}}_{{{\boldsymbol{\hat{y}}} \sim p_{{\boldsymbol{\hat{y}}} | {\boldsymbol{x}} }}}}\left[ {d\left( {{\boldsymbol{x}},{\boldsymbol{\hat{x}}}} \right)} \right],
	\end{equation}
	where ${\bm{\phi}}$ and ${\bm{\theta}}$ encapsulate all the network parameters of $f_e$ and $f_d$.

	To address the needs of single adaptive scenarios, either SNR or rate, we further propose two schemes: SNR adaptive SwinJSCC and Rate adaptive SwinJSCC, as illustrated in Fig. \ref{Fig6}(b) and \ref{Fig6}(c). These schemes are individually optimized for a single conditional change (i.e., channel state and target rate) in a communication scenario, improving performance. In comparison to the scheme incorporating the Rate ModNet module, the SNR adaptive SwinJSCC necessitates an additional FC layer to fine-tune the channel number of latent representations to achieve the target CBR $R$. Both the proposed Channel ModNet and Rate ModNet modules are plug-in modules, enabling effective working in scenarios requiring only single adaptive. Although their robustness is inferior to the scheme adapting to both SNR and rate simultaneously, there is a marginal enhancement in end-to-end transmission performance.

	\section{Adaptive Channel-Dependent Mechanism}\label{adaptive mechanism}
	
	To enhance the transmission quality and reconstructed image fidelity by adapting to real-time channel conditions, we propose two key plug-in modules, namely Channel ModNet and Rate ModNet. The Channel ModNet dynamically adjusts the parameters and configuration of the model to optimally adapt to varying channel qualities by modeling and predicting the input SNR. The Rate ModNet utilizes masks to rescale the output features to dynamically select the appropriate channel bandwidth rate to maximize transmission efficiency. These two adaptive mechanisms aim to achieve optimal transmission performance across diverse SNR conditions and target rates, thereby improving system robustness and reliability.

	\subsection{Channel ModNet}
	
	\begin{figure*}[t]
		\setlength{\abovecaptionskip}{0.cm}
		\setlength{\belowcaptionskip}{-0.cm}
		\centering{\includegraphics[scale=0.72]{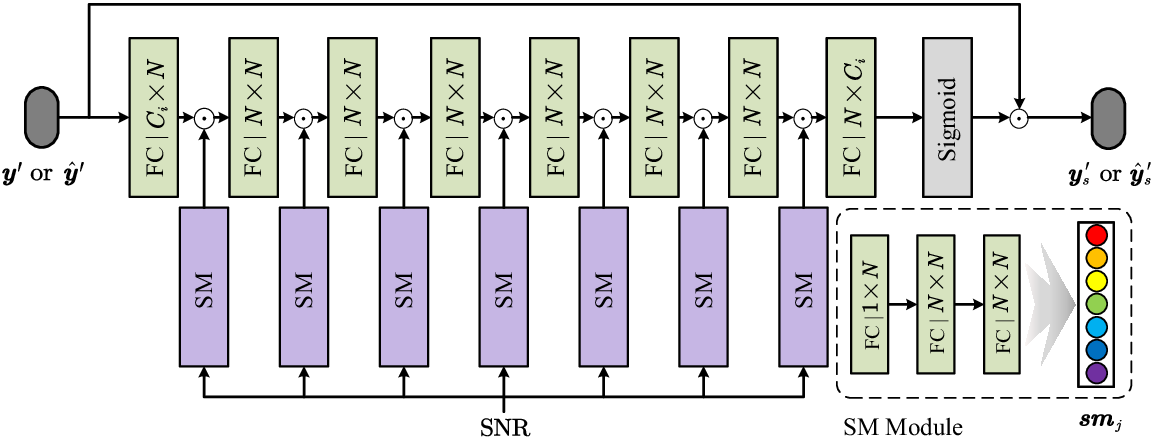}}
		\caption{The architecture of Channel ModNet. $C_i$ and $N$ denote the number of channels in $\boldsymbol{y}^\prime$ or $\boldsymbol{\hat{y}}^\prime$ and the number of intermediates of FC in $\boldsymbol{sm}_j$ respectively. }\label{Fig7}
	\end{figure*}
	
	In this paper, we design an adaptive channel-dependent mechanism to enable our end-to-end image transmission system to automatically adapt to the changes in channel state without relying on gradient descent. We propose the ``Channel ModNet'' as a plug-in module for the SwinJSCC scheme, which modulates the output of several Transformer stages. The Channel ModNet is inserted in both the encoder and decoder and modulates the intermediate tokens based on the instantaneous wireless channel state. Under different channel states, different resource allocation strategies should be adopted to implicitly adjust the source and channel coding rates inside the SwinJSCC encoder and decoder, achieving higher-quality transmission and reconstruction images.
	
	As illustrated in Fig. \ref{Fig5}, the semantic feature map $\boldsymbol{y}^\prime$ is fed into the Channel ModNet for modulation, which considers the channel state information. The resulting modulated embeddings are subsequently fed into the Rate ModNet to obtain the semantic feature map $\boldsymbol{y}$. At the receiver, the symbol $\boldsymbol{\hat{y}}$ undergoes modulation by our Channel ModNet to restore the semantic feature embeddings $\boldsymbol{\hat{y}}^\prime$. As such, our Channel ModNet is integrated into the encoder and decoder to facilitate modulating the intermediate tokens according to the instant wireless channel state.
	
	The Channel ModNet comprises two key elements, SNR modulation (SM) and FC layers, for the features. As illustrated in Fig. \ref{Fig7}, the Channel ModNet includes $8$ FC layers, interspersed with $7$ SM module. The SM module is a three-layered FC network with the channel SNR input. It transforms the channel state value $\text{SNR}$ into an $N$-dimensional vector $\boldsymbol{sm}_j$. The multiple SM modules are cascaded sequentially in a coarse-to-fine manner. The previously modulated features are then fed into subsequent SM modules, allowing for the achievement of arbitrary target modulation by assigning a corresponding SNR value. The mapping procedures from $\text{SNR}$ to $\boldsymbol{sm}_j$ are
	\begin{subequations}
		\begin{equation}\label{eq_5a}
			\boldsymbol{sm}^{(1)}_j = \text{ReLU}(\boldsymbol{W}^{(1)} \cdot \text{SNR} + \boldsymbol{b}^{(1)}), 
		\end{equation}
		\begin{equation}\label{eq_5b}
			\boldsymbol{sm}^{(2)}_j = \text{ReLU}(\boldsymbol{W}^{(2)} \cdot \boldsymbol{sm}^{(1)}_j + \boldsymbol{b}^{(2)}), 
		\end{equation}
		\begin{equation}\label{eq_5c}
			\boldsymbol{sm}_j = \text{Sigmoid}(\boldsymbol{W}^{(3)} \cdot \boldsymbol{sm}^{(2)}_j + \boldsymbol{b}^{(3)}),         
		\end{equation}
	\end{subequations}
	where Sigmoid is the activation function, ReLU denotes the rectified linear unit activation function, $\boldsymbol{W}^{(k)}$ and $\boldsymbol{b}^{(k)}$ are the affine function parameters, and their corresponding bias. 
	
	Therefore, the channel state information is associated with the $N$-dimensional tensor $\boldsymbol{sm}_j$. Subsequently, the input feature will be fused with $\boldsymbol{sm}_j$ in the element-wise product,i.e.,
	\begin{equation}\label{eq_6}
		\bm{output} = \boldsymbol{input} \odot \boldsymbol{sm}_j 
	\end{equation}
	Here, $\boldsymbol{input}$ denotes the feature output from the previous FC layer, and $\boldsymbol{output}$ feeds into the next FC layer. Multiple SM modules are cascaded sequentially in a coarse-to-fine manner in Fig. \ref{Fig7}. The previous modulated features are fed into subsequent SM modules. Finally, the decisions generated by the fully connected layer are broadcasted into the same spatial size as $\boldsymbol{y}^\prime$. In our design, different $\boldsymbol{sm}_j$ pay attention to channel states such that SNR modulation is comprehensively considered in a channel-wise attention fashion.
	
	\subsection{Rate ModNet}
	
	\begin{figure*}[t]
		\setlength{\abovecaptionskip}{0.cm}
		\setlength{\belowcaptionskip}{-0.cm}
		\centering{\includegraphics[scale=0.72]{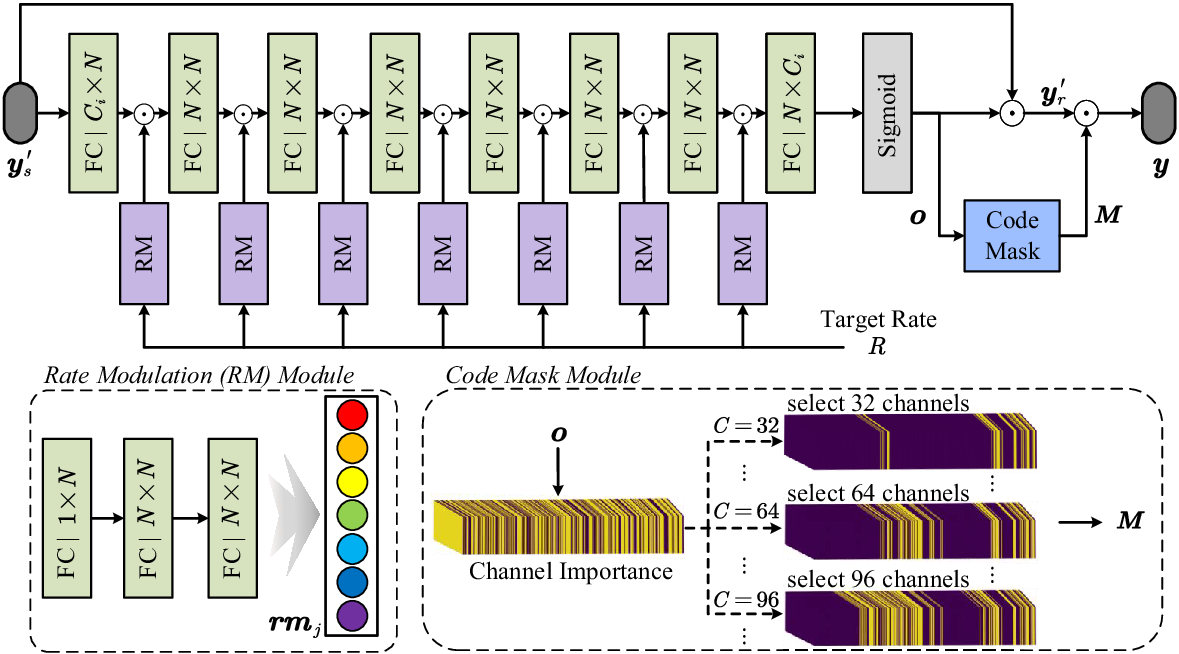}}
		\caption{The architecture of Rate ModNet. For different channel numbers $C_i$, the code mask module will sort according to the channel importance and select the most important $C_i$ channels. The highlighted part in the importance ranking indicates that the channel is more important.}\label{Fig8}
	\end{figure*}
	
	For practical end-to-end image transmission, both the channel states and the target rate are known to be continuously variable, making it necessary to develop a method to adapt to these changes in real time. To address this issue, we propose using Rate ModNet as a plug-in module to enhance the model's performance and facilitate automatic adaptation to any target rate. This module is intended to rescale the previously proposed Channel ModNet output, as shown in Fig. \ref{Fig5}. We obtain different neural-syntax for different target rates $R$ to generate a more precise codec function. In this way, our proposed schemes are well-suited for achieving efficient and reliable image transmission in wireless communication systems and improved robustness in dynamic wireless transmission. 
	
	As demonstrated in Fig. \ref{Fig8}, the Rate ModNet has $8$ FC layers separated by $7$ rate modulation (RM) modules and a code mask module. The RM module is a three-layered FC network that takes the target rate $R$ as input and converts it into an $N$-dimensional vector $\boldsymbol{rm}_j$. The mapping process from $R$ to $\boldsymbol{rm}_j$  is identical to that of $\text{SNR}$ to $\boldsymbol{sm}_j$ employed in the Channel ModNet. This design enables the proposed model to automatically adjust to various target rates in real time, thereby enhancing the performance of the codec function. Afterward, the input feature is fused with $\boldsymbol{rm}_j$ through an element-wise product.
	
	Notably, some of the components in Rate ModNet are similar to those in Channel ModNet, with rate modulation (RM) sharing the same structure as the SM module. Modifying the latent feature $\boldsymbol{y}^\prime$ using the learned $\boldsymbol{rm}_j$ to adapt to the target rate is insufficient. Thus, we proposed a novel code mask module to analyze the relevance of the rate representation $\boldsymbol{o}$ and rank it based on the channel dimension. Here, the relevance is determined by averaging the spatial dimension values of the latent representation. Following the ranking, we choose the top $C$ dimensions from the relevance ranking, which is calculated for a given transmission rate. Subsequently, a binary vector mask $\boldsymbol{M}$ is generated based on this selection process, which contains $C$ ones, with the remaining elements being zeros. The mask $\boldsymbol{M}$ is then applied to the corresponding modulated embeddings $\boldsymbol{y}_r^\prime$, resulting in a channel input symbol vector $\boldsymbol{y} = \boldsymbol{y}_s^\prime \odot \boldsymbol{M}$. This approach enhances the adaptability of the proposed model to the target rate by selectively rescaling the relevant modulated features.

	In our proposed, the mask  $\boldsymbol{M}$ determines the number of channels we choose and the position of each channel in the latent representation. Therefore, the decoding process relies on accurately receiving side information masks $\boldsymbol{M}$, which play a critical role. To ensure the transmission of $\boldsymbol{M}$ without loss, we use entropy coding, which consumes additional bandwidth. This additional bandwidth is negligible for high-resolution images compared to the bandwidth required for transmitting feature values. In contrast, for low-resolution images, the size of the transmitted side information is almost the same as that of the feature maps, leading to a substantial loss in performance. Thus, we do not consider transmitting low-resolution images using rate adaptive scheme.

	\section{Experimental Results}\label{Experimental}
	
	\subsection{Experimental Setup}
	
	\subsubsection{Datasets}
	
	Our SwinJSCC model is trained using the DIV2K dataset\cite{agustsson2017ntire}. During training, images are randomly cropped into patches with dimensions of $256 \times 256$. We evaluate the performance of SwinJSCC using both the Kodak dataset\cite{Kodak} at the size of $512 \times 768$ and the CLIC2021 testset\cite{CLIC2021} with approximate 2K resolution images. For a fair comparison, all images are cropped to multiples of 128 to avoid padding for neural codecs. We also carried out some experiments on low-resolution images to further validate the model performance. We use the CIFAR10 dataset\cite{CIFAR10} for training and testing the SNR adaptive SwinJSCC models. 
	
	\subsubsection{Comparison Schemes}
	
	We compare our proposed SwinJSCC scheme with the CNN-based deep JSCC scheme\cite{bourtsoulatze2019deep}, DeepJSCC-V scheme\cite{zhang2023predictive}, and classical separation-based source and channel coding schemes. Specifically, we employ the BPG codec \cite{BPG} for compression combined with 5G LDPC codes\cite{LDPC_5G} for channel coding. Here, we considered 5G LDPC codes with a block length of 6144 bits for different coding rates and quadrature amplitude modulations (QAM). Moreover, the ideal capacity-achieving channel code is also considered during the evaluation. Apart from these, we also compare our improved single adaptive SwinJSCC with the deep JSCC scheme and the base SwinJSCC for end-to-end transmission. For simplicity, we mark our versatile SwinJSCC model with both SNR and rate adaptation as ``SwinJSCC w/ SA\&RA'', the SwinJSCC with only SNR adaptation as ``SwinJSCC w/ SA'', the SwinJSCC with only rate adaptation is labeled as ``SwinJSCC w/ RA'', and the baseline SwinJSCC is labeled as ``SwinJSCC w/o SA\&RA'', where ``SA'' represents SNR adaptive and ``RA'' stands for rate adaptive.
	
	\begin{figure*}[htbp]
		\setlength{\abovecaptionskip}{0.cm}
		\setlength{\belowcaptionskip}{-0.cm}
		\begin{center}
			\hspace{-.1in}
			\subfigure[]{
				\includegraphics[width=0.33\linewidth]{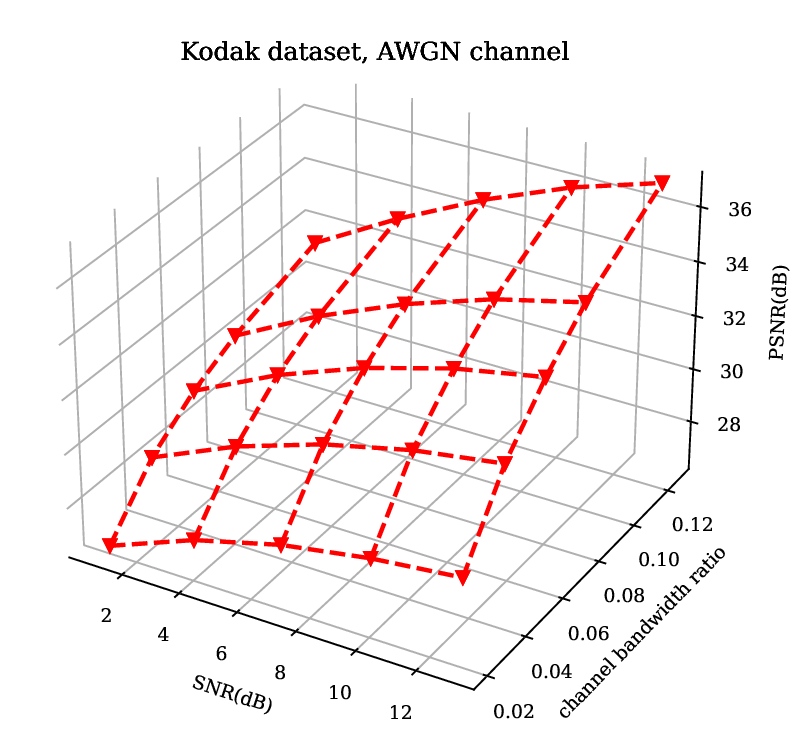}}
			\hspace{-.1in}
			\subfigure[]{
				\includegraphics[width=0.33\linewidth]{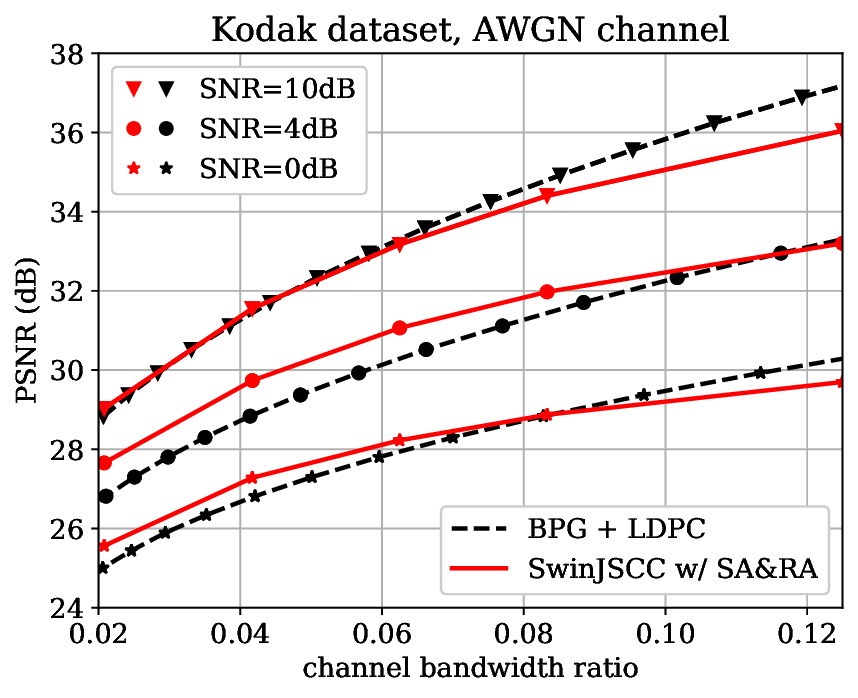}}
			\hspace{-.1in}
			\subfigure[]{
				\includegraphics[width=0.33\linewidth]{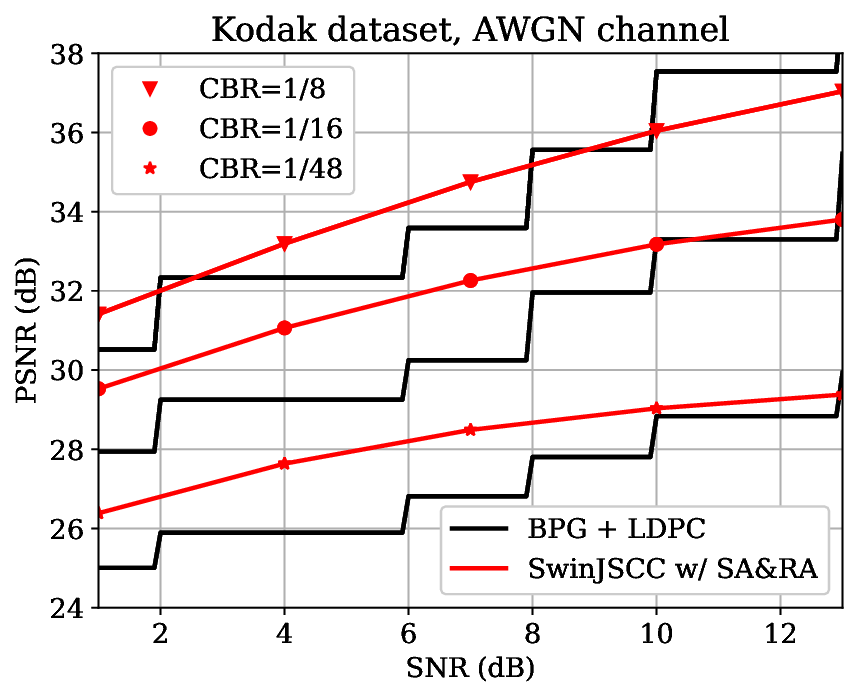}}
			
			\hspace{-.1in}
			\subfigure[]{
				\includegraphics[width=0.33\linewidth]{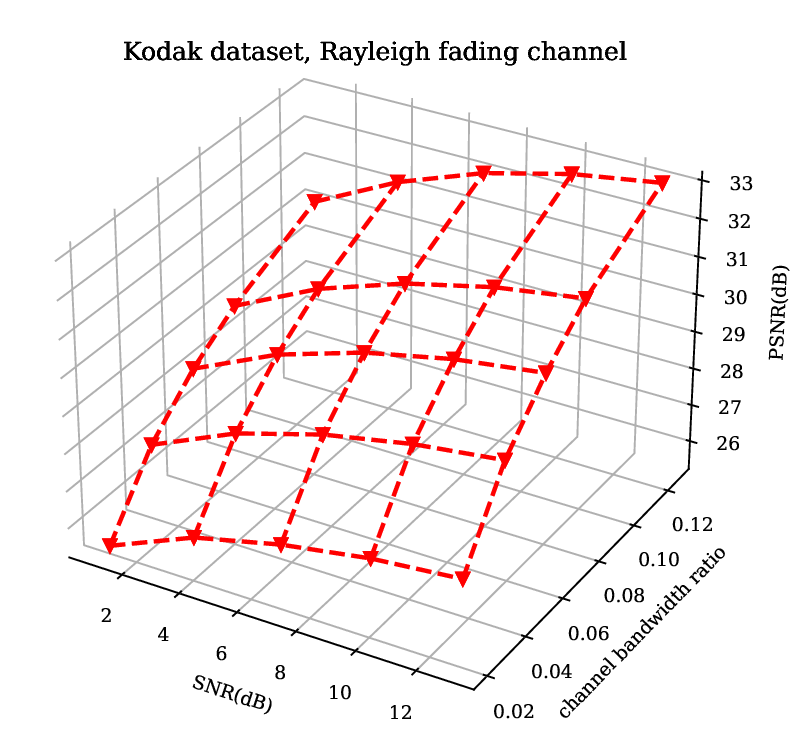}}
			\hspace{-.1in}
			\subfigure[]{
				\includegraphics[width=0.33\linewidth]{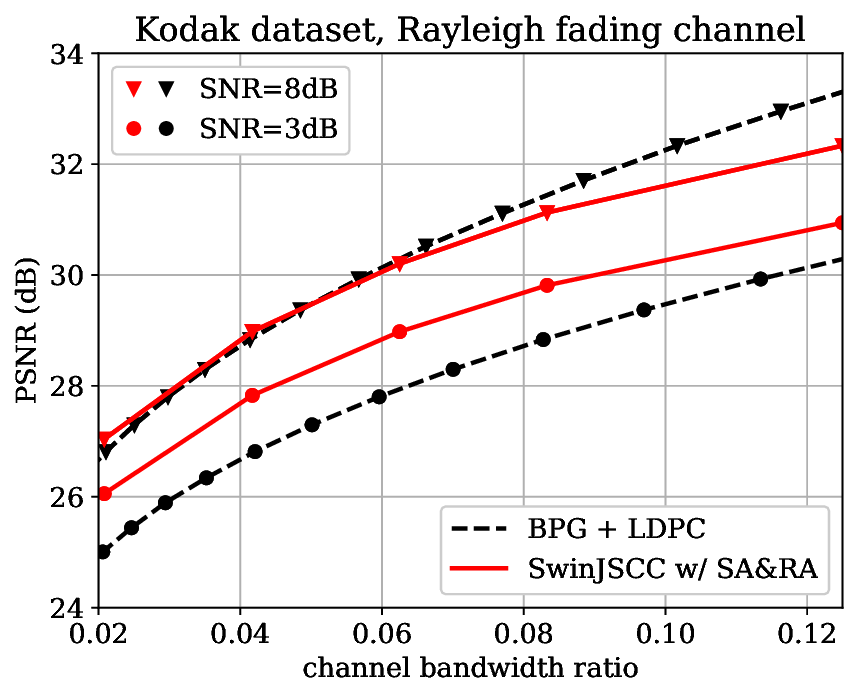}}
			\hspace{-.1in}
			\subfigure[]{
				\includegraphics[width=0.33\linewidth]{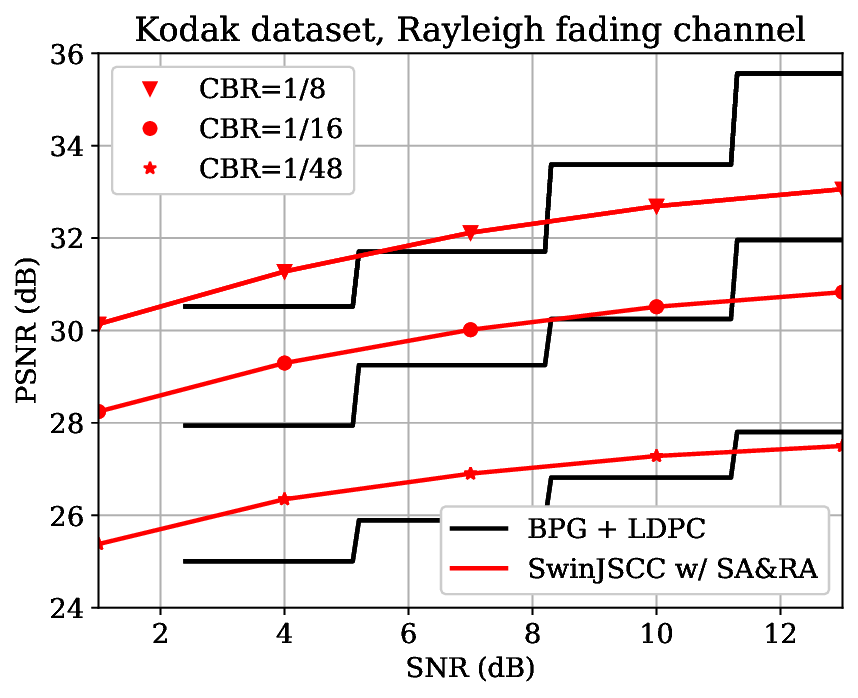}}
			\caption{SNR-rate-distortion comparison over AWGN and fast Rayleigh fading channel on Kodak dataset. (a)(d) show the SNR-rate-PSNR mesh obtained by our SwinJSCC model. (b) compares RD curves of different coded transmission schemes at SNR = 0dB, 4dB, and 10dB. (c) compares SNR-PSNR curves under the CBR constraint CBR = 1/48, 1/16, 1/8. (e) compares RD curves of different coded transmission schemes at SNR = 3dB and 8dB. (f) compares SNR-PSNR curves under the CBR constraint CBR = 1/48, 1/16, 1/8.}\label{Fig9}  
		\end{center}
	\end{figure*}

	\subsubsection{Evaluation Metrics}
	
	We qualify the end-to-end image transmission performance of the proposed SwinJSCC models and other comparison schemes using the widely used pixel-wise metric PSNR and the perceptual metric MS-SSIM\cite{wang2003multiscale}. For PSNR, we optimized our model by the mean square error (MSE) loss function between $\boldsymbol{x}$ and $\boldsymbol{\hat{x}}$. For MS-SSIM, the loss function $d$ is set as 1 $-$ MS-SSIM. It is usually known that a higher PSNR/MS-SSIM indicates a better performance.

	\subsubsection{Model Training Details}
	
	The number of stages in SwinJSCC varies with training image resolution. For low-resolution images, we use 2 stages with $[N_1, N_2]=[2, 4]$, $[C_1, C_2]=[128, 256]$, and the window size is set to 2. For large-resolution images, we use 4 stages $[N_1, N_2, N_3, N_4]=[2, 2, 6, 2]$, $[C_1, C_2, C_3, C_4]=[128, 192, 256, 320]$, and the window size is set to 8. For training the SwinJSCC model, We first train the whole model with a fixed rate ($R=0.125$) and channel state ($\text{SNR}=13$dB). Then, we only change the given rate ($R=[0.0208, 0.0417, 0.0625, 0.0833, 0.125]$) to train the whole model. Finally, we train the whole model with a variable rate and variable channel state ($\text{SNR}=[1, 4, 7, 10, 13]$dB) to obtain a universal wireless image transmission model. For training SNR adaptive model and rate adaptive model, we first train other parameters except for the Channel or Rate ModNet over the wireless channel. Then, the whole proposed model is trained with Channel or Rate ModNet. 
	
	We exploit the Adam optimizer with a learning rate of $1 \times 10^{-4}$, and the batch size is set to 128 and 16 for the CIFAR10 dataset and DIV2K dataset, respectively. The SwinJSCC model is trained under the channel with a uniform distribution of SNR$_{train}$ from 1dB to 13 dB and a target rate $R=[0.0208, 0.0417, 0.0625, 0.0833, 0.125]$. All implementations were done on Pytorch, and it takes about four days to train each step model using a single RTX 3090 GPU for the DIV2K dataset.

	\begin{figure*}[t]
		\setlength{\abovecaptionskip}{0.cm}
		\setlength{\belowcaptionskip}{-0.cm}
		\begin{center}
			\subfigure[]{
				\hspace{-0.1in}
				\includegraphics[width=0.33\linewidth]{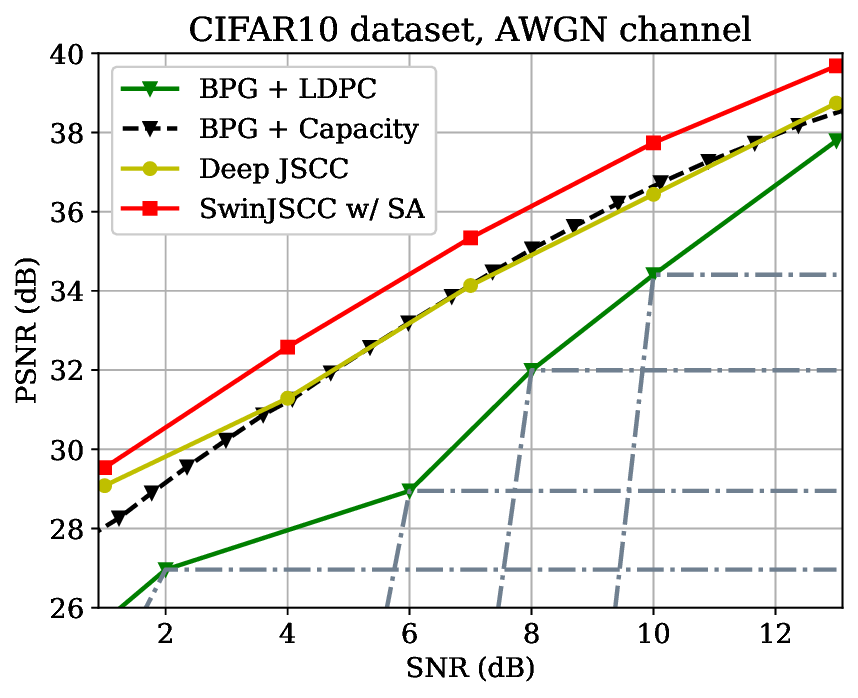}}
			\hspace{-0.1in}
			\subfigure[]{
				\includegraphics[width=0.33\linewidth]{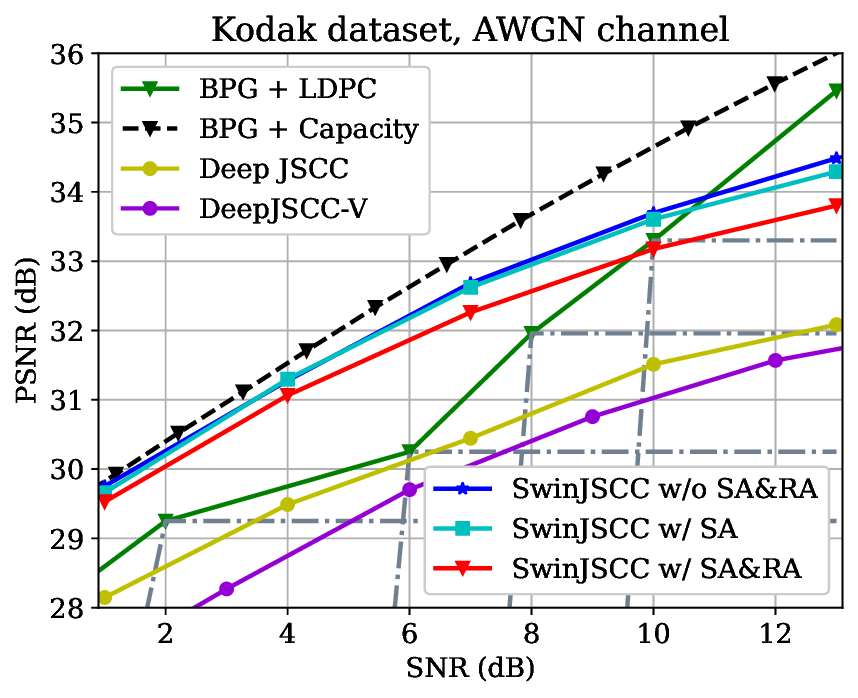}}
			\hspace{-0.1in}
			\subfigure[]{
				\includegraphics[width=0.33\linewidth]{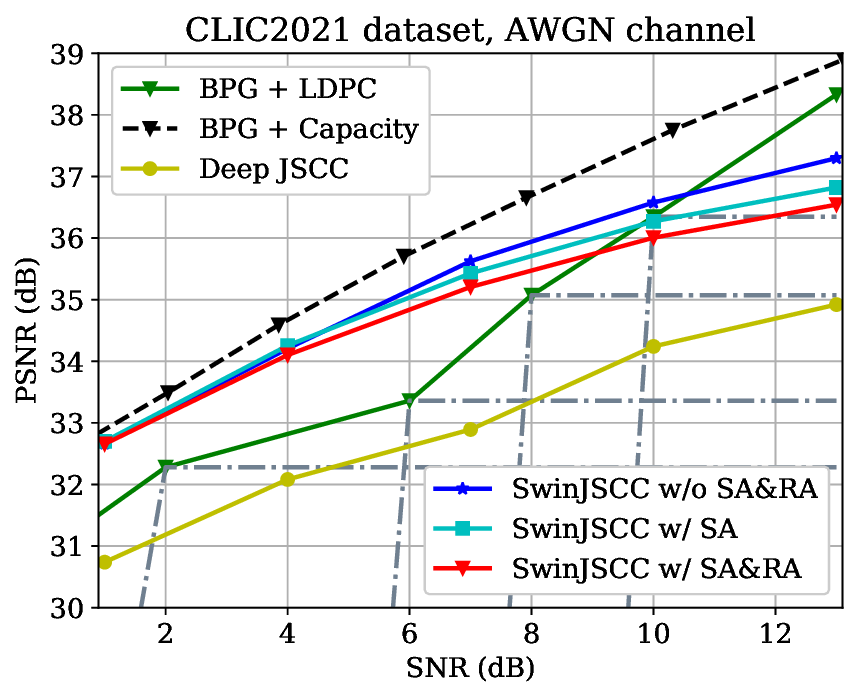}}
			
			\hspace{-0.1in}
			\subfigure[]{
				\includegraphics[width=0.33\linewidth]{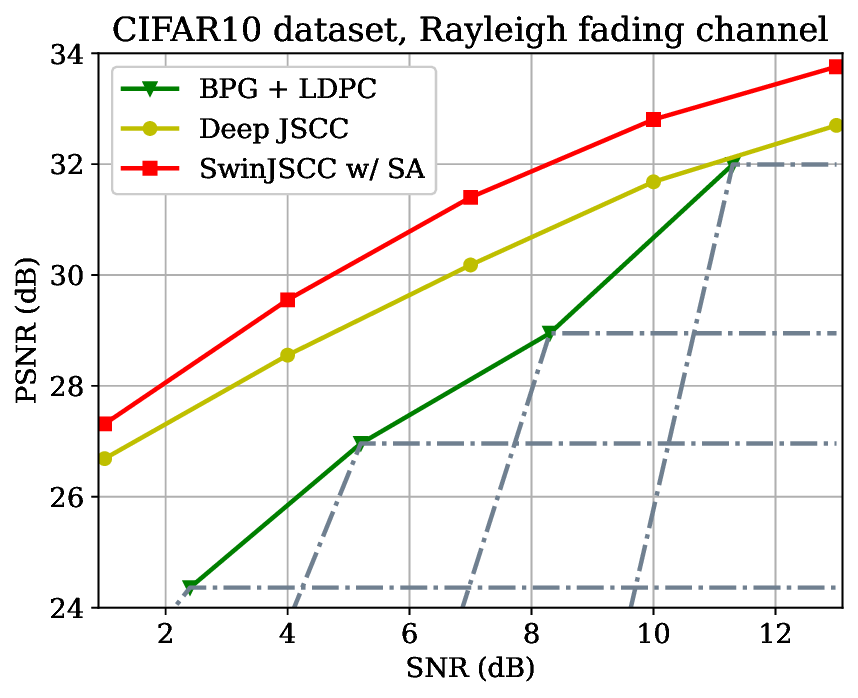}}
			\hspace{-0.1in}
			\subfigure[]{
				\includegraphics[width=0.33\linewidth]{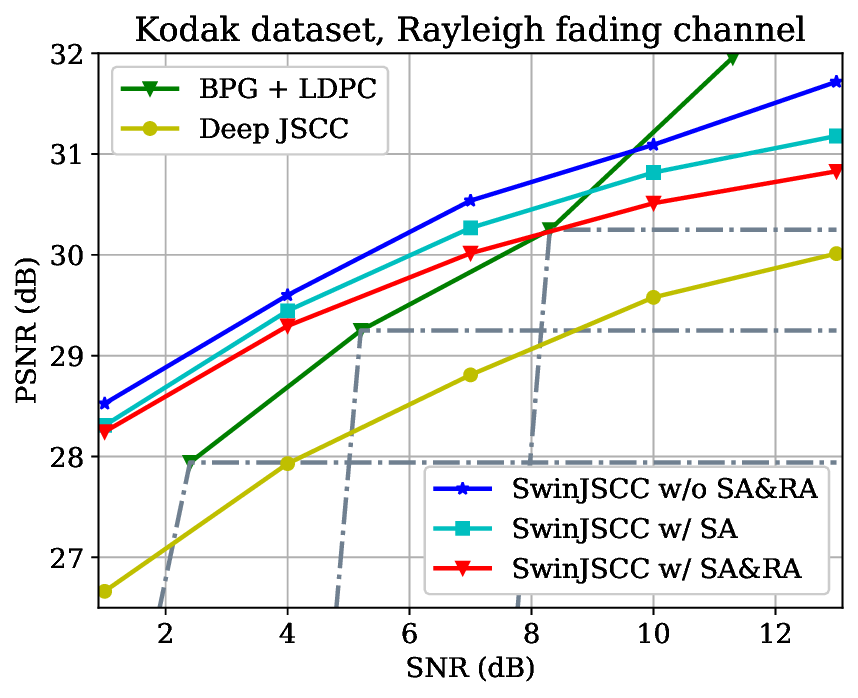}}
			\hspace{-0.1in}
			\subfigure[]{
				\includegraphics[width=0.33\linewidth]{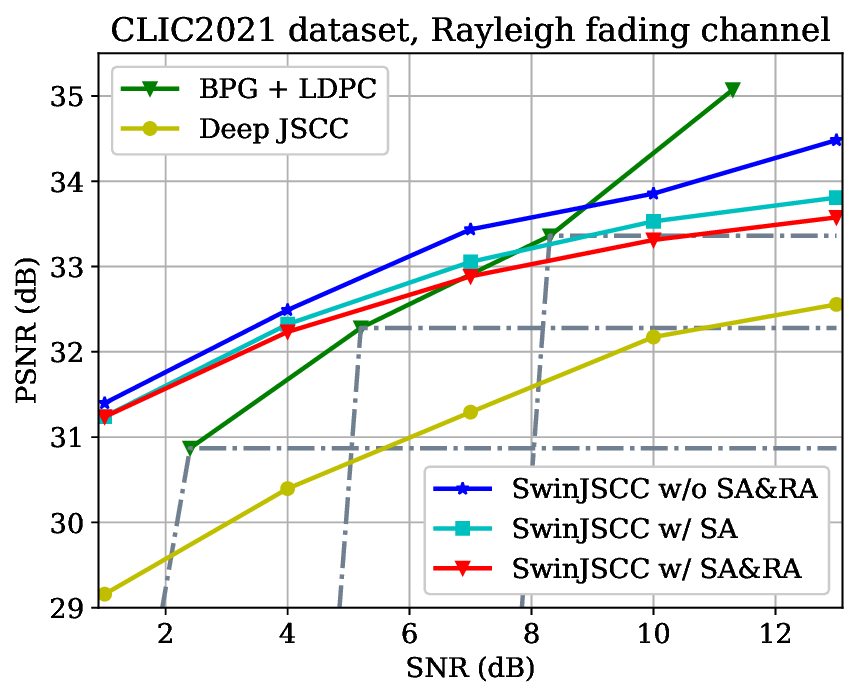}}
			
			\caption{(a)$\sim$(c) PSNR performance versus the SNR over the AWGN channel. (d)$\sim$(f) PSNR performance versus the SNR over the fast Rayleigh fading channel. The average CBR is set to 1/3, 1/16, and 1/16 for the CIFAR10 dataset, Kodak dataset, and CLIC21 dataset.}\label{Fig10}  
		\end{center}
	\end{figure*}

	\begin{figure*}[t]
		\setlength{\abovecaptionskip}{0.cm}
		\setlength{\belowcaptionskip}{-0.cm}
		\begin{center}
			\hspace{-0.1in}
			\subfigure[]{
				\includegraphics[width=0.33\linewidth]{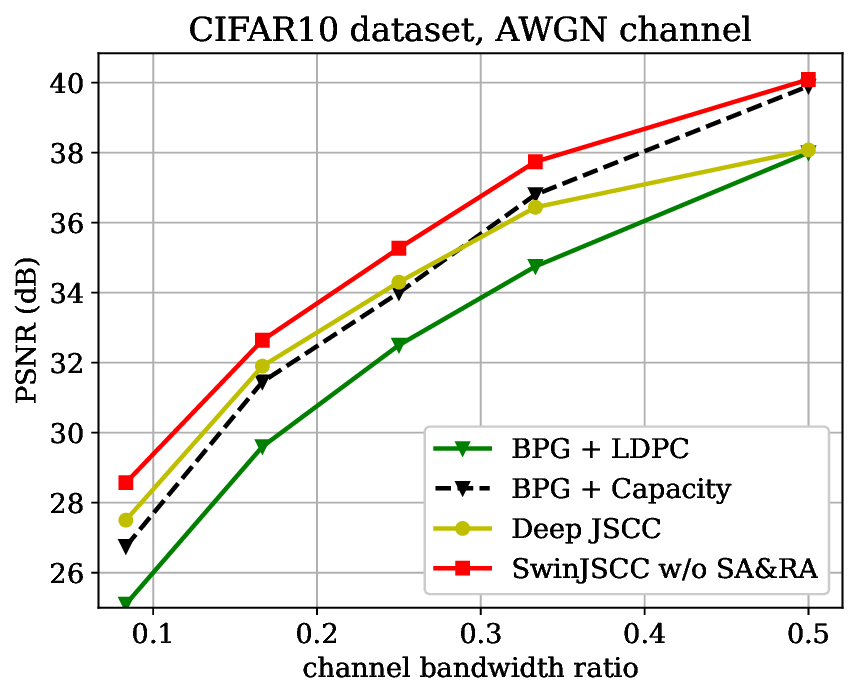}}
			\hspace{-0.1in}
			\subfigure[]{
				\includegraphics[width=0.33\linewidth]{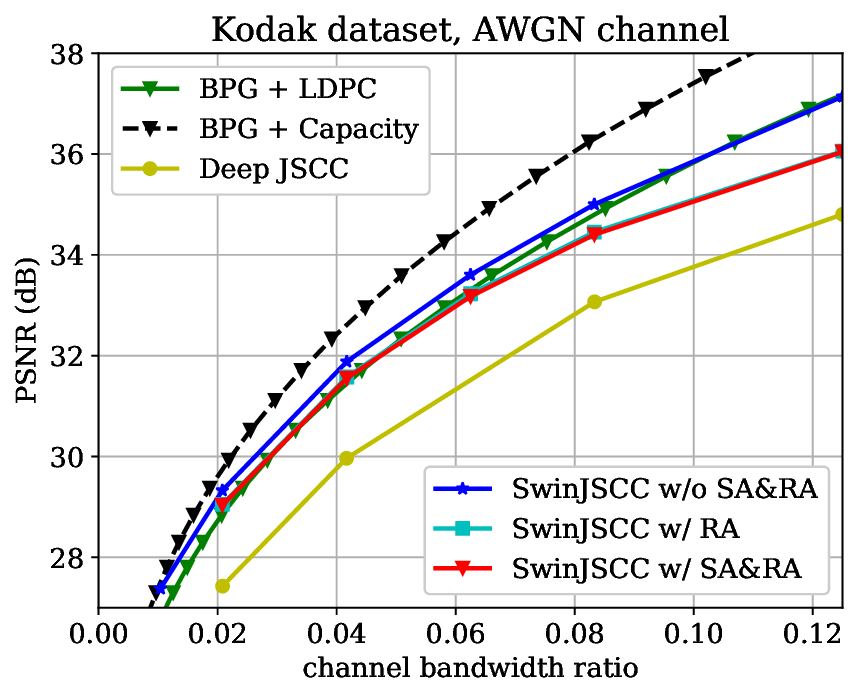}}
			\hspace{-0.1in}
			\subfigure[]{
				\includegraphics[width=0.33\linewidth]{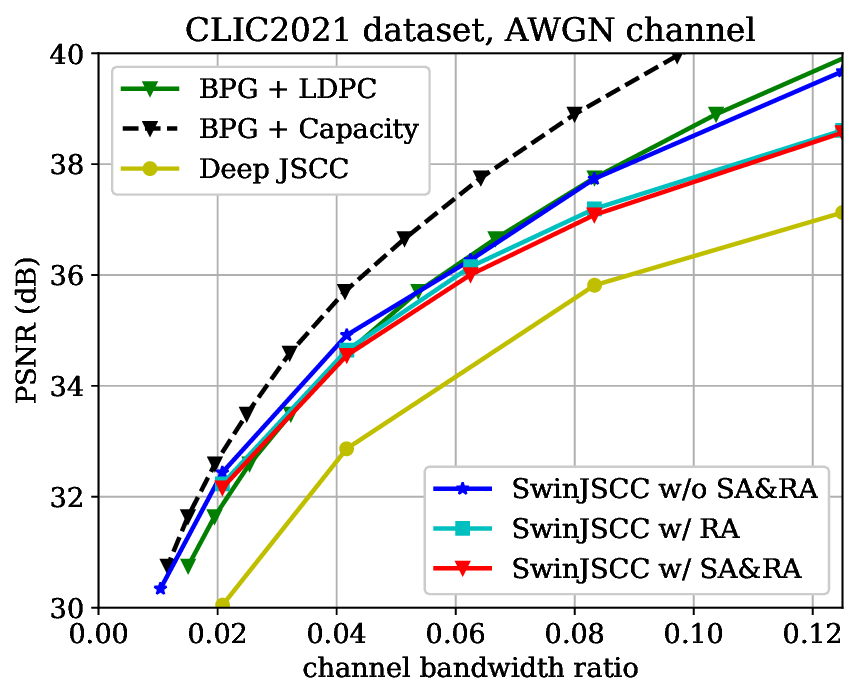}}
			
			\hspace{-0.1in}
			\subfigure[]{
				\includegraphics[width=0.33\linewidth]{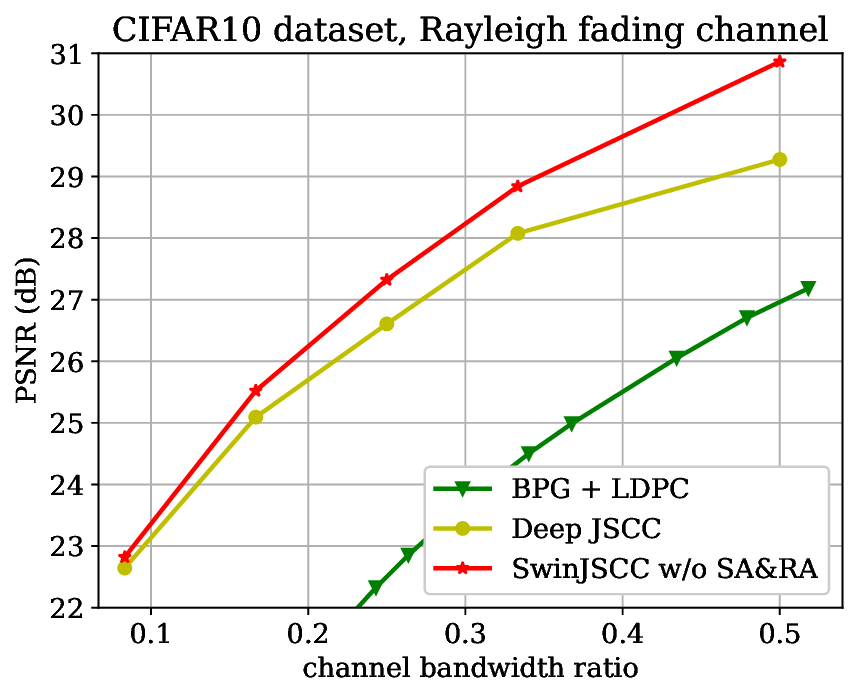}}
			\hspace{-0.1in}
			\subfigure[]{
				\includegraphics[width=0.33\linewidth]{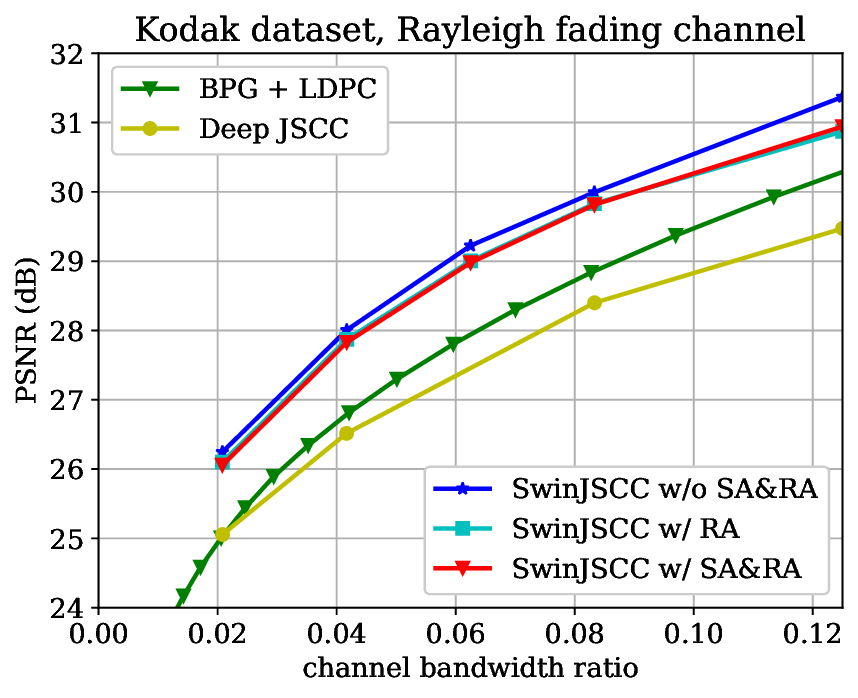}}
			\hspace{-0.1in}
			\subfigure[]{
				\includegraphics[width=0.33\linewidth]{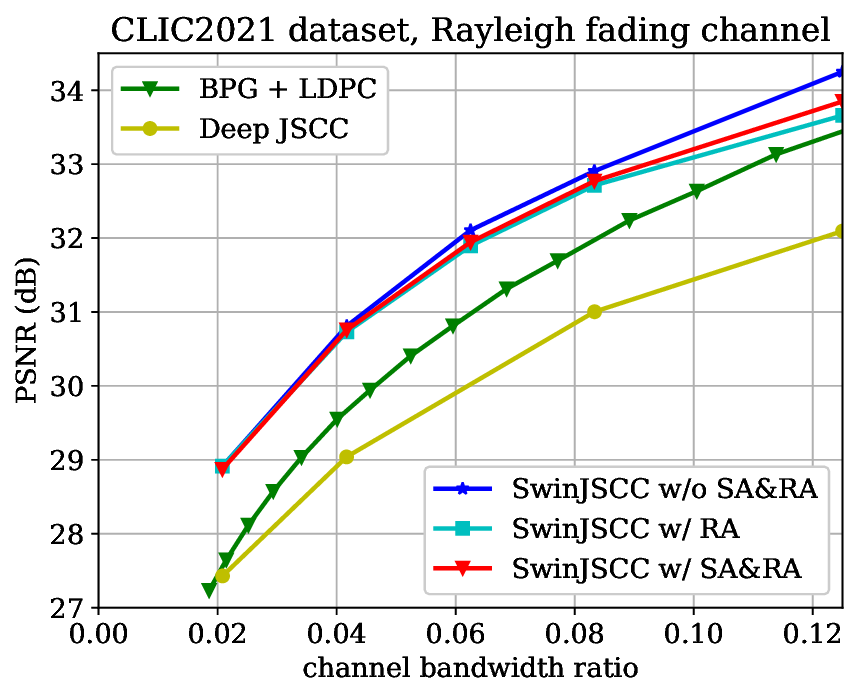}}
			\caption{(a)$\sim$(c) PSNR performance versus the CBR over the AWGN channel at $\text{SNR} = 10\text{dB}$. (d)$\sim$(f) PSNR performance versus the CBR over the fast Rayleigh fading channel at $\text{SNR} = 3\text{dB}$. }\label{Fig11}  
		\end{center}
	\end{figure*}
	
	\begin{figure*}[t]
		\setlength{\abovecaptionskip}{0.cm}
		\setlength{\belowcaptionskip}{-0.cm}
		\begin{center}
			\hspace{-0.1in}
			\subfigure[]{
				\includegraphics[width=0.33\linewidth]{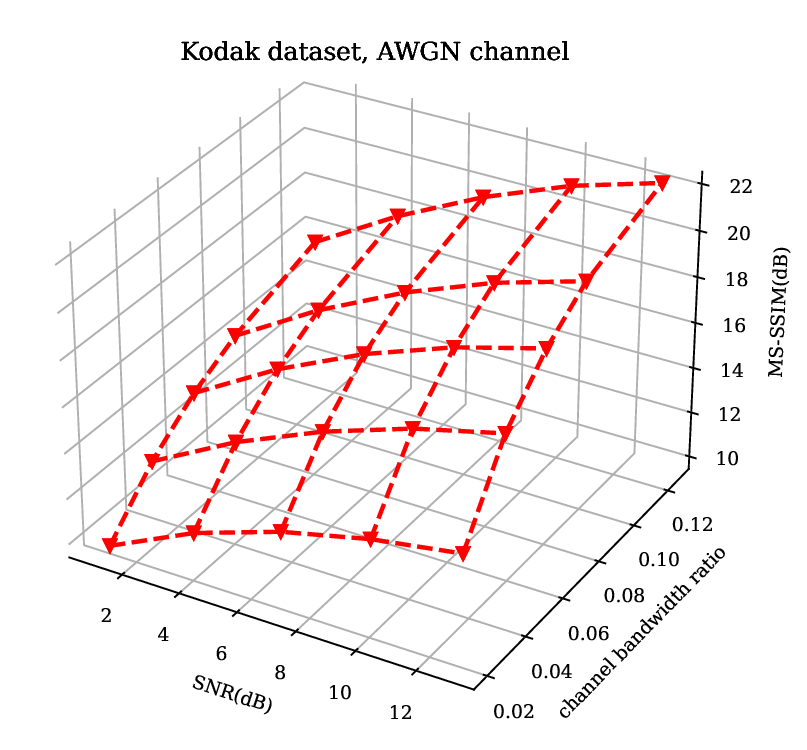}}
			\hspace{-0.1in}
			\subfigure[]{
				\includegraphics[width=0.33\linewidth]{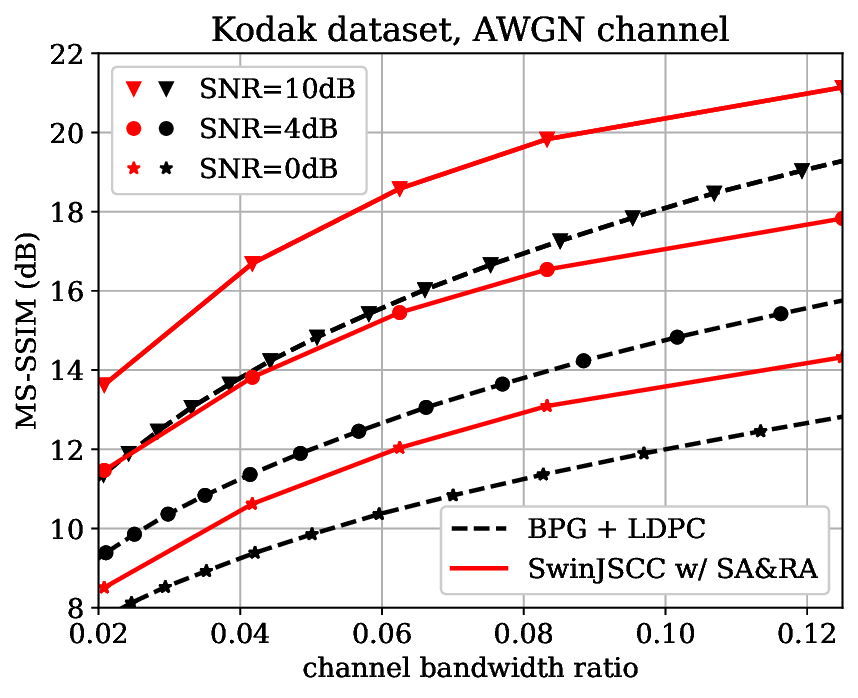}}
			\hspace{-0.1in}
			\subfigure[]{
				\includegraphics[width=0.33\linewidth]{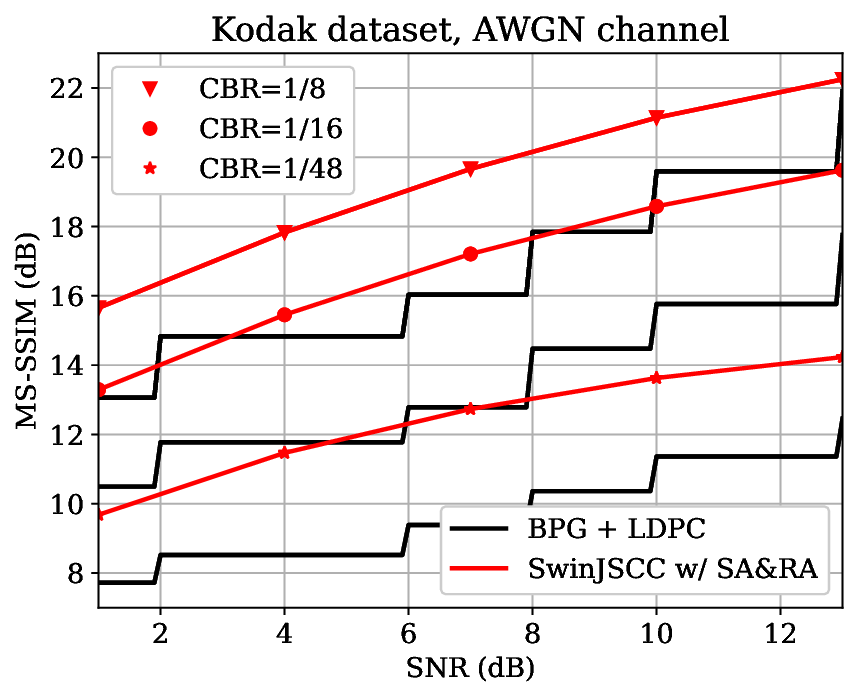}}
			\caption{SNR-rate-MS-SSIM comparison over AWGN and channel on Kodak dataset. (a) show the SNR-rate-MS-SSIM mesh obtained by our SwinJSCC w/ SA\&RA model. (b) compares RD curves of different coded transmission schemes at SNR = 0dB, 4dB, and 10dB. (c) compares SNR-MS-SSIM curves under the CBR constraint CBR = 1/48, 1/16, 1/8.}\label{Fig12}  
		\end{center}
	\end{figure*}

	\begin{figure*}[t]
		\setlength{\abovecaptionskip}{0.cm}
		\setlength{\belowcaptionskip}{-0.cm}
		\begin{center}
			\hspace{-.1in}
			\subfigure[]{
				\includegraphics[width=0.24\linewidth]{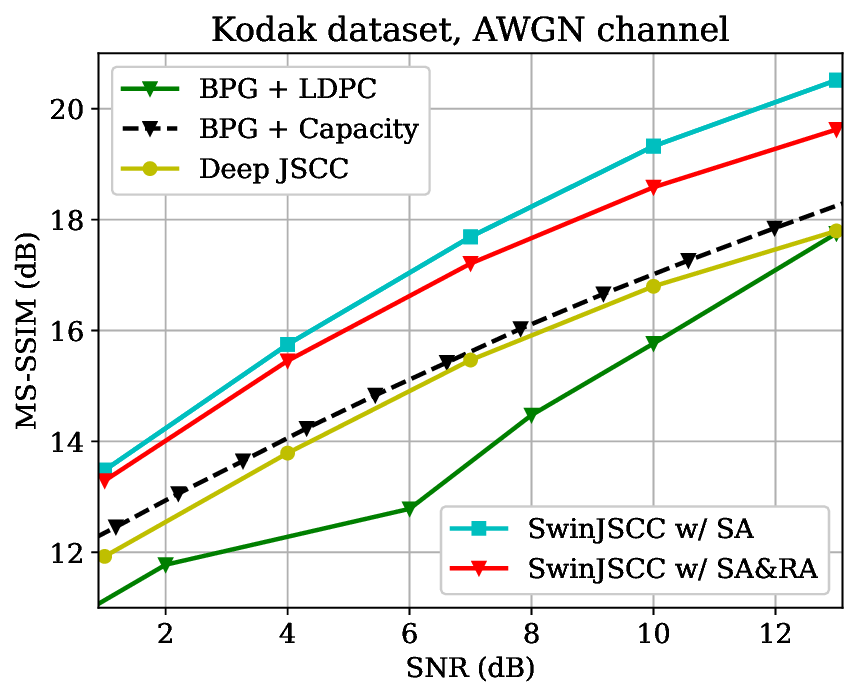}}
			\hspace{-.1in}
			\subfigure[]{
				\includegraphics[width=0.24\linewidth]{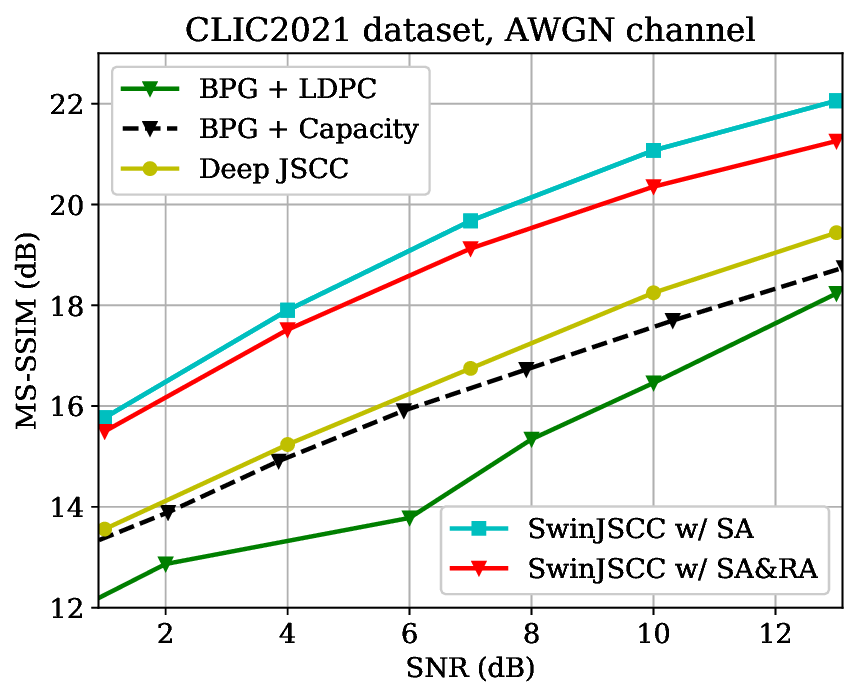}}
			\hspace{-.1in}
			\subfigure[]{
				\includegraphics[width=0.24\linewidth]{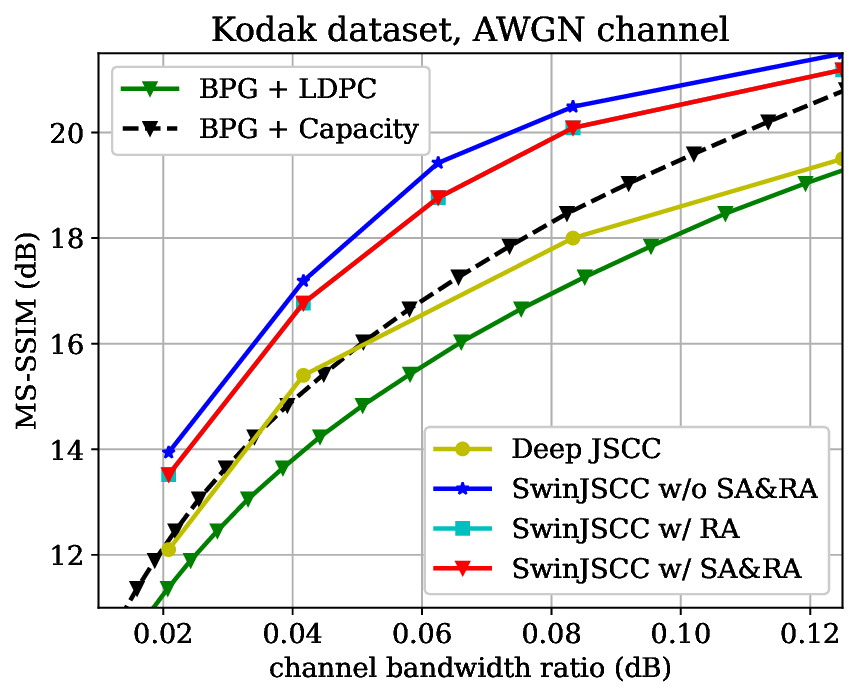}}
			\hspace{-.1in}
			\subfigure[]{
				\includegraphics[width=0.24\linewidth]{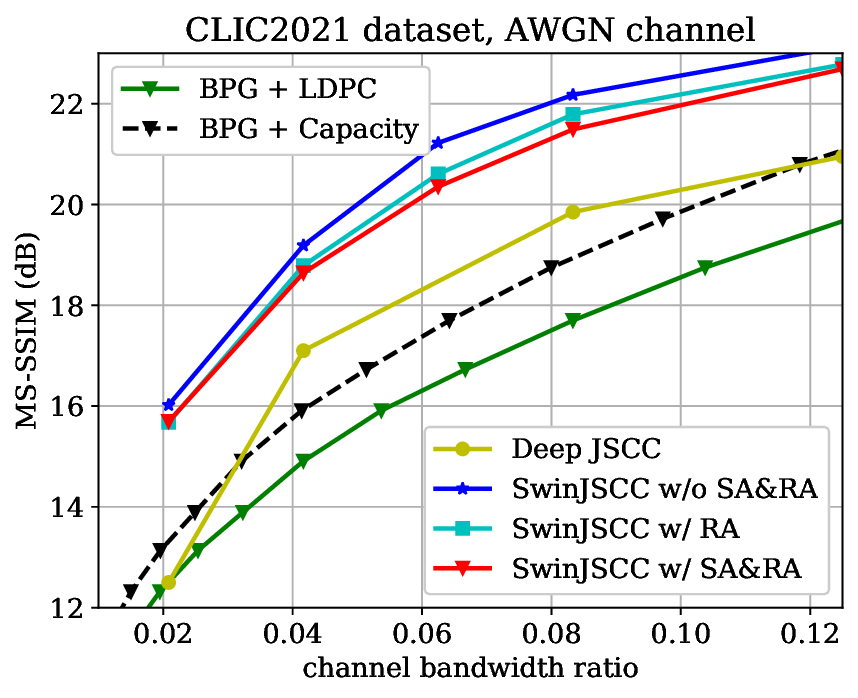}}
			\caption{(a)$\sim$(b) MS-SSIM performance versus the SNR over the AWGN channel and the average CBR is set to 1/16. (c)$\sim$(d) MS-SSIM performance versus the CBR over the AWGN channel at $\text{SNR} = 10\text{dB}$.}\label{Fig13}  
		\end{center}
	\end{figure*}
	
	\subsection{Results Analysis}
	
	\subsubsection{PSNR Performance}
	
	Fig. \ref{Fig9} depicts the performance of the proposed SwinJSCC w/ SA\&RA model under different SNR values and different CBR constraints over the AWGN channel and Rayleigh fading channel. Notably, each surface point is generated from the same SwinJSCC w/ SA\&RA model. Our proposed model demonstrates strong adaptability to varying channel conditions with different SNRs and CBRs, resulting in comparable or superior performance to ``BPG + LDPC''. Results indicate that our SwinJSCC w/ SA\&RA as a universal model can achieve satisfactory continuous rate and SNR adaptation in a single model with negligible performance loss.
	
	Furthermore, Fig. \ref{Fig10} shows the PSNR performance versus the SNR over the AWGN and Rayleigh fading channels. For the SwinJSCC w/o SA\&RA, each point in the curve is obtained from a separate training model. Data extraction from the paper \cite{zhang2023predictive} informs the DeepJSCC-V model, an adaptive wireless image compression and transmission scheme. For the ``BPG + LDPC'' scheme, we choose the best-performing configuration of coding rate and modulation (the green dashed lines) based on the adaptive modulation and coding (AMC) standard \cite{3gpp} under each specific SNR and plot the envelope. 
	
	Fig. \ref{Fig11} shows the PSNR performance versus the CBR over the AWGN and Rayleigh fading channels. For the ``BPG + LDPC'' scheme, we employ a 2/3 rate (4096, 6144) LDPC code with 16-ary quadrature amplitude modulation (16QAM). The SwinJSCC w/ RA model requires side information $\boldsymbol{M}$ to assist decoding. However, for low-resolution datasets such as CIFAR10, the cost of transmitting side information is exceedingly high. Consequently, we employ the SwinJSCC w/o SA\&RA model to evaluate the performance of the CIFAR10 dataset. For high-resolution datasets, the cost of transmitting side information is minimal on the order of $10^{-4}$ and can be considered negligible.
	
	\begin{figure*}[t]
		\setlength{\abovecaptionskip}{0.cm}
		\setlength{\belowcaptionskip}{-0.cm}
		\begin{center}
		\centering{\includegraphics[scale=0.6]{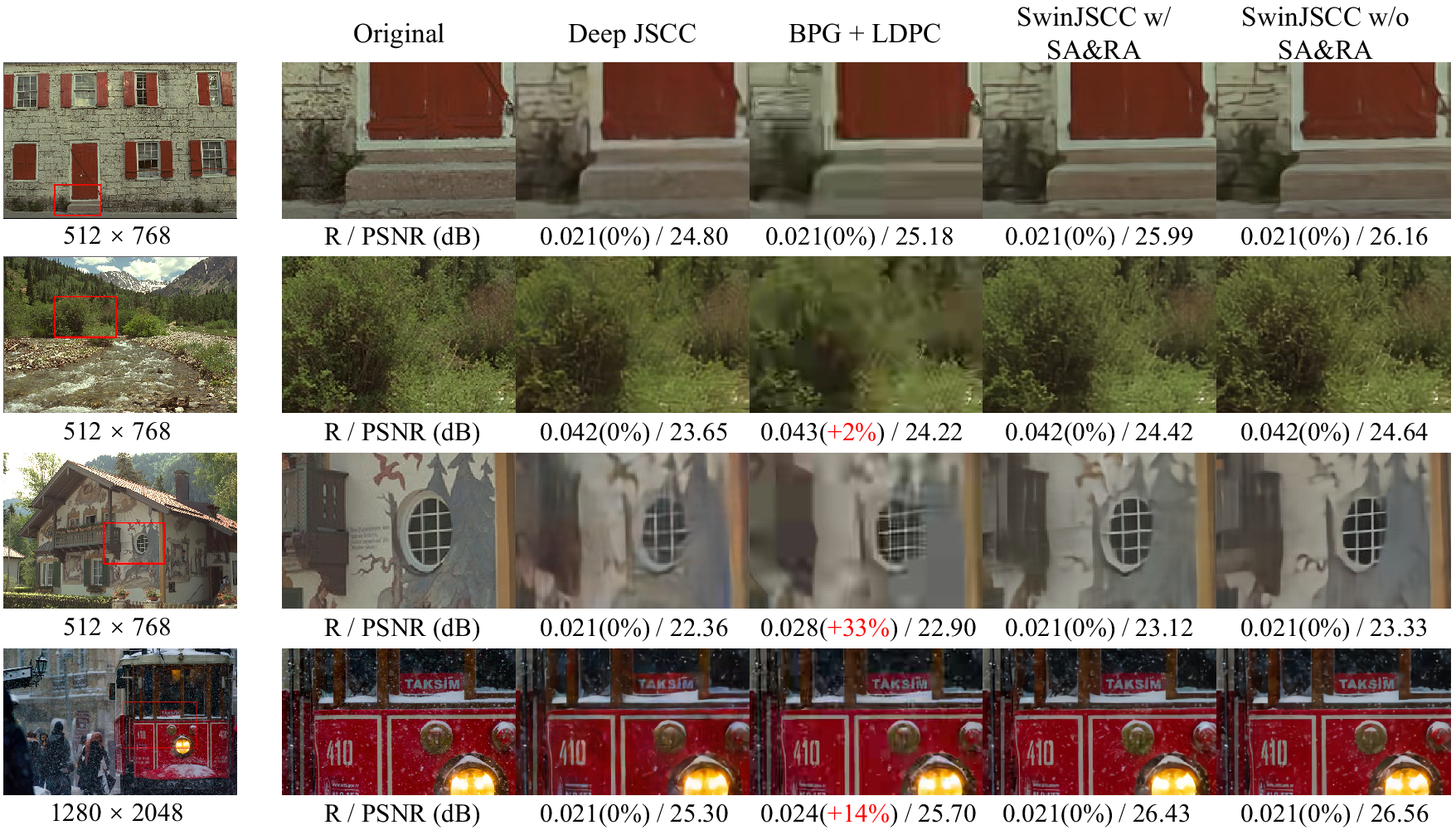}}			
			\caption{The first two rows are examples of visual comparison under AWGN channel at $\text{SNR} = 10$dB. The last two rows are examples of visual comparison under AWGN channel at $\text{SNR} = 3$dB. The first, second, and third to sixth columns show the original image, original patch, and reconstructions of different transmission schemes, respectively. The red and blue numbers indicate the percentage of bandwidth cost increase and saving compared to deep JSCC.}\label{Fig14}
		\end{center}
	\end{figure*}

	We observe that our proposed SwinJSCC w/ SA\&RA model outperforms the CNN-based deep JSCC scheme and DeepJSCC-V scheme for all SNRs, and its performance gap widens with the increase of image resolution due to the enhanced model capacity by incorporating Transformers. For the CIFAR10 dataset, our model and deep JSCC scheme significantly outperform the ``BPG + LDPC'' and ``BPG + Capacity''. However, for high-resolution images, the performance of CNN-based deep JSCC degrades significantly and falls behind the separation-based scheme. Our proposed models maintain considerable performance regarding classical separation-based schemes, particularly in the low SNR regions. Besides, as observed in Fig. \ref{Fig10}, the proposed models also achieve graceful degradation as deep JSCC does when the testing SNR decreases from the training SNR. At the same time, the performance of the separation-based ``BPG + LDPC'' transmission scheme reduces drastically (known as the cliff effect).	
	
	Upon examination of the proposed models, the SwinJSCC w/o SA\&RA model was found to exhibit the optimal performance, followed by the SwinJSCC w/ SA and SwinJSCC w/ RA models, while the performance of the SwinJSCC w/ SA\&RA model was comparatively inferior. We hypothesize that the semantic feature has been resized by the Channel ModNet, leading to the loss of some features on channels, which is the underlying reason for this performance difference. Despite the slight performance loss, the SwinJSCC w/ SA\&RA model's ability to dynamically adapt to varying channel and communication conditions provides a significant advantage in practical wireless image transmission scenarios. 
	
	\subsubsection{MS-SSIM Performance}
	
	To provide a more comprehensive assessment of our proposed model, we conducted further experiments to evaluate its performance using the multi-scale structural similarity index (MS-SSIM) metric. MS-SSIM is a multi-scale perceptual metric that approximates human visual perception well, and its values are between 0 and 1. Since most MS-SSIM values obtained in our experiments are higher than 0.9, we converted them in dB to improve the legibility, using the formula MS-SSIM(dB) = $-10\log(1-\text{MS-SSIM})$. 
	
	Fig. \ref{Fig12} depicts the performance of the proposed SwinJSCC w/ SA\&RA model under different SNR values and different CBR constraints over the AWGN channel. Our proposed SwinJSCC w/ SA\&RA model exhibits strong adaptability to diverse channel conditions, achieving comparable or superior MS-SSIM performance to "BPG + LDPC" while enabling satisfactory continuous rate and SNR adaptation in a single model without significant performance degradation.		
	
	Fig. \ref{Fig13} shows the MS-SSIM performance over the AWGN channel. Results reveal that the proposed models can outperform competitors by a significant margin, improving more on high-resolution images and high CBR regions. Compared to the PSNR results, we observed that the learning-based schemes outperform the BPG series because BPG compression is designed to optimize for squared error with hand-crafted constraints. Furthermore, the SwinJSCC w/o SA\&RA model performs better than the SwinJSCC w/ RA and SwinJSCC w/ SA\&RA models.
	
	\begin{table*}[ht]
		\renewcommand{\arraystretch}{1.3}
		\centering
		\small
		
		\caption{Averaged encoding/decoding latency on the Kodak dataset.}  
		\label{table1} 
		
		\begin{tabular}{m{4cm}<{\raggedright}|c|c|c|c|c}
			
			\Xhline{1pt}
			
			\multicolumn{1}{c|}{\multirow{2}{*}{Transmission scheme}} &\multicolumn{1}{c|}{\multirow{2}{*}{BD-rate (\%)}} & \multicolumn{1}{c|}{\multirow{2}{*}{Parameters (M)}} &\multicolumn{1}{c|}{\multirow{2}{*}{Inference time}} & \multicolumn{2}{c}{End-to-end latency}\\ \cline{5-6} 
			\multirow{2}{*}{}&\multirow{2}{*}{} &\multirow{2}{*}{} &\multirow{2}{*}{} & Encoding & Decoding \tabularnewline
			
			\hline
			\hline

			BPG + LDPC\cite{BPG,LDPC_5G} & \centering{0} &  \centering{--} & \centering{$>$7.6s} & \centering{$>$670ms} & \centering{$>$7.3s} \tabularnewline
			\hline
			
			ADJSCC\cite{xu2021wireless} & \centering{36.03} &  \centering{14.66} & \centering{212ms} & \centering{67.3ms} & \centering{94ms} \tabularnewline
			
			\hline
			
			SwinJSCC w/o SA\&RA & \centering{--29.71} &  \centering{18.34} & \centering{151ms} & \centering{13ms} & \centering{13ms} \tabularnewline      
			
			SwinJSCC w/ RA & \centering{--25.78} &  \centering{18.34 + {4.87}} & \centering{167ms} & \centering{35ms} & \centering{12ms} \tabularnewline

			SwinJSCC w/ SA & \centering{--26.40} &  \centering{18.34 + {9.86}} & \centering{177ms} & \centering{35ms} & \centering{16ms} \tabularnewline

			SwinJSCC w/ SA\&RA & \centering{--25.10} &  \centering{18.34 + {14.73}} & \centering{191ms} & \centering{38ms} & \centering{16ms}\tabularnewline
			
			\hline
			
			\Xhline{1pt}
			
		\end{tabular}
		\label{Table1}
	\end{table*}
	
	\subsubsection{Visualization Results}
	
	To further demonstrate the effectiveness of our proposed models, we provide a set of visually intuitive results on the testset as shown in Fig. \ref{Fig14}. The visual comparisons are conducted under both AWGN and Rayleigh fading channels, revealing the robustness and adaptivity of our proposed models. From these results, we can observe that our proposed SwinJSCC w/ SA\&RA transmission scheme exhibits much better visual quality with lower channel bandwidth cost. In particular, it avoids artifacts effectively and produces a high-fidelity reconstruction with more generated details, while the traditional ``BPG + LDPC'' scheme exhibits blocking artifacts. Thus, it can better support the human vision demands in semantic communications.
	
	We evaluate the end-to-end processing latency of these coded transmission schemes on the Rayleigh fading channel and show the metrics in Table \ref{Table1}, including the BD-rate, single model parameters, inference time, and latency. The experiment is implemented on PyTorch 1.9.1 with an Inter Xeon Gold 6226R CPU and one RTX 3090 GPU. We conducted ten trials on the Kodak dataset with a batch size of 1 to obtain the average encoding and decoding time per image, which allowed us to calculate the encoding/decoding latency and inference time. It can be seen that our SwinJSCC series schemes run much faster than the classical scheme ``BPG + LDPC'', mainly due to the absence of LDPC coding time and saving more than 20.57$\%$ channel bandwidth cost. Despite its larger model size, our proposed can provide better performance and run faster than the ADJSCC scheme. Compared with the SwinJSCC w/o SA\&RA, the encoding/decoding time of our improved version increases. However, it is still valuable since the overall adaptability of the model has been significantly improved, making it more suitable for practical wireless transmission scenarios.

	\begin{figure}[t]
		\setlength{\abovecaptionskip}{0.cm}
		\setlength{\belowcaptionskip}{-0.cm}
		\centering{\includegraphics[scale=0.37]{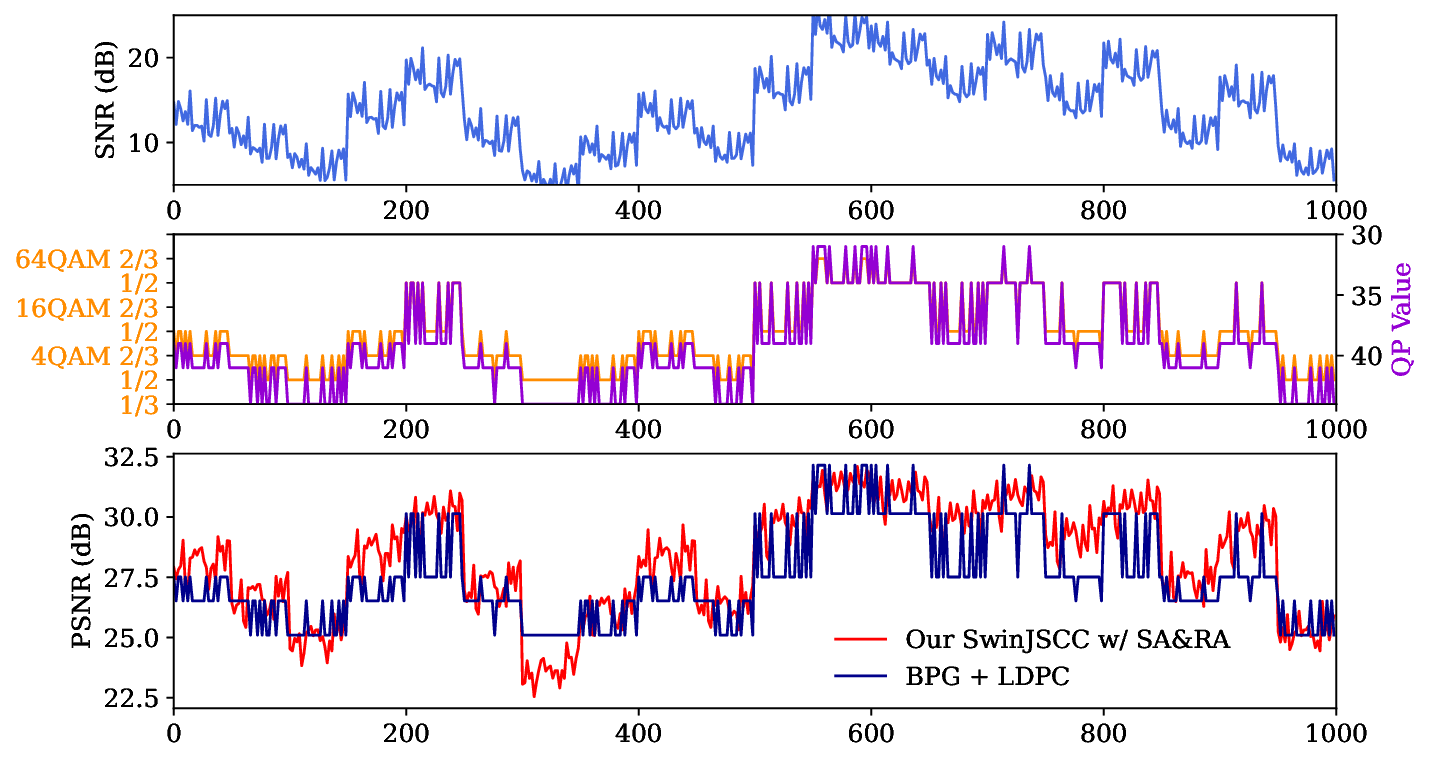}}
		\caption{Comparisons of the image quality between the ``BPG + LDPC'' and our SwinJSCC w/ SA\&RA under a practical multipath fading channel\cite{yang2022ofdm}, where the fading coefficient varies with the frame number. The top subfigure shows the instant channel SNR, and the middle subfigure shows the adaptive coded modulation scheme and the quantization parameter (QP) in ``BPG + LDPC''. The bottom subfigure plots the PSNR value of each frame under the CBR = 0.0625.}\label{Fig15}
	\end{figure}
	
	Furthermore, Fig. \ref{Fig15} shows the performance under a practical multipath fading channel. To investigate the model's transient performance in realistic channel settings, we conducted a transmission experiment using the Kodim14 image from the Kodak dataset. Specifically, we transmitted each image one thousand times and analyzed the results presented in Fig. \ref{Fig15}. In comparison to the traditional ``BPG + LDPC'' scheme that utilizes a layered design with a limited number of quantization parameters (QPs) in source compression and channel-coded modulation schemes, our proposed SwinJSCC w/ SA\&RA model with response networks exhibits greater flexibility and coding gain, enabling it to react more sensitively to SNR variations and outperform traditional schemes.
	
	\begin{figure*}[ht]
		\setlength{\abovecaptionskip}{0.cm}
		\setlength{\belowcaptionskip}{-0.cm}
		\begin{center}
			\hspace{-.1in}
			\subfigure[(a)]{
				\includegraphics[width=0.34\linewidth]{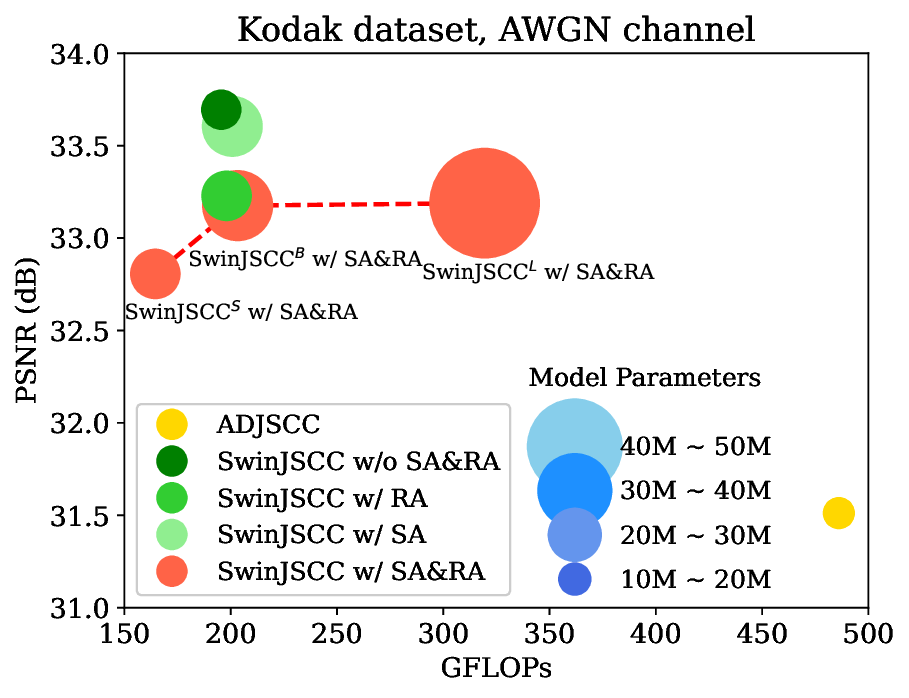}}
			\hspace{-.1in}
			\subfigure[(b)]{
				\includegraphics[width=0.32\linewidth]{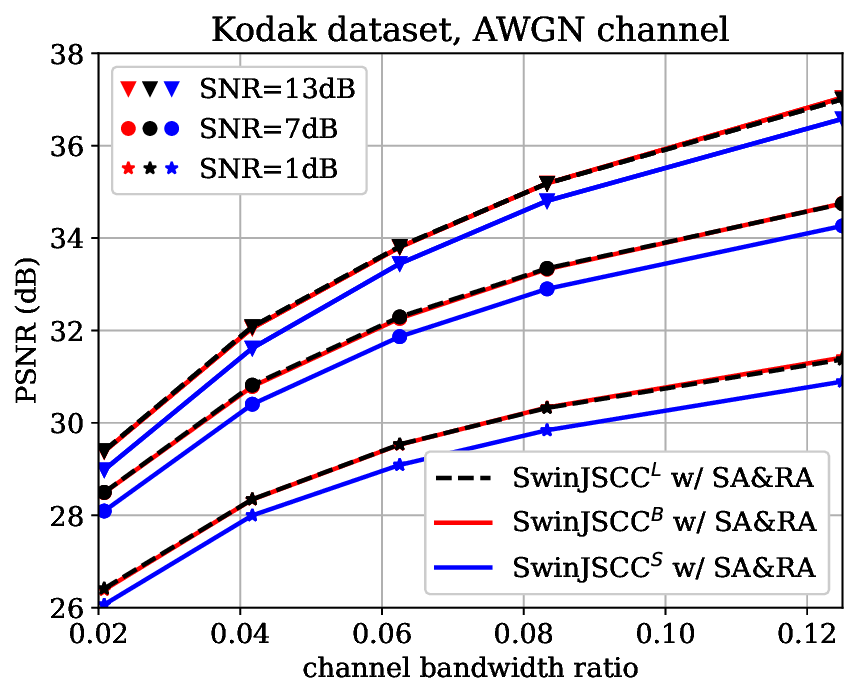}}
			\hspace{-.1in}
			\subfigure[(c)]{
				\includegraphics[width=0.32\linewidth]{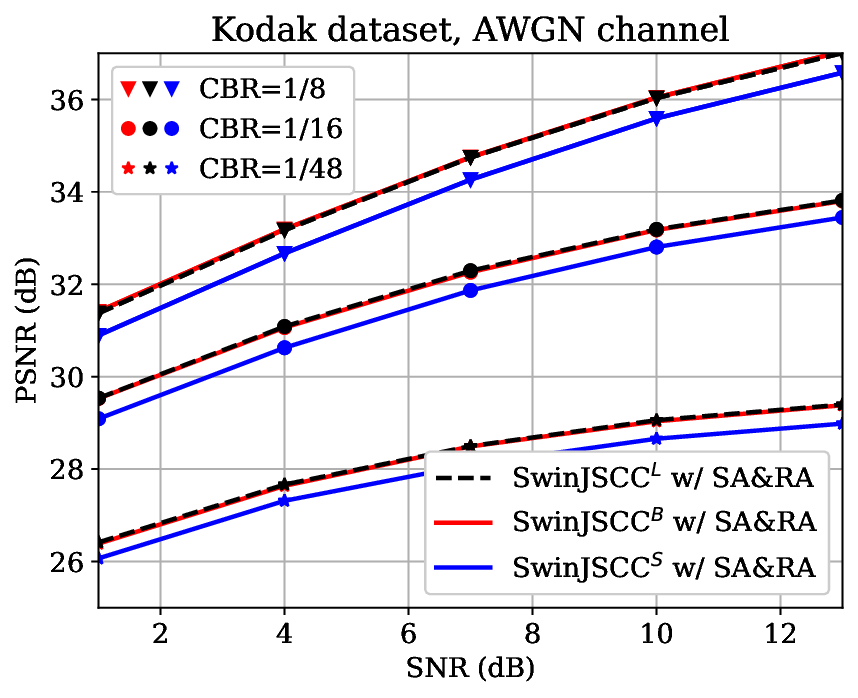}}
			\caption{(a) PSNR performance versus compute, with bubble size representing the number of parameters. (b) compares RD curves of different coded transmission schemes at SNR = 1dB, 7dB, and 13dB. (c) compares SNR-PSNR curves under the CBR constraint CBR = 1/48, 1/16, 1/8.}\label{Fig16}  
		\end{center}
	\end{figure*}
	
	\subsubsection{Ablation Study}
	We build our base model setting the Swin Transformer layer numbers as $[N_1, N_2, N_3, N_4]=[2, 2, 6, 2]$, called SwinJSCC$^{B}$ w/ SA\&RA. To investigate the impact of model size and computational complexity on performance, we introduce two additional small-size and large-size variants: SwinJSCC$^{S}$ w/ SA\&RA and SwinJSCC$^{L}$ w/ SA\&RA which are versions of about 0.75x and 1.5x the model size and computational complexity, respectively. The architecture parameters of these model variants are:
	\begin{itemize}
		\item[$\bullet$]  SwinJSCC$^{S}$ w/ SA\&RA:\\ layer numbers $[N_1, N_2, N_3, N_4]=[2, 2, 2, 2]$ 
		\item[$\bullet$]  SwinJSCC$^{L}$ w/ SA\&RA:\\ layer numbers $[N_1, N_2, N_3, N_4]=[2, 2, 18, 2]$
	\end{itemize}
	
	As shown in Fig. \ref{Fig16}, our proposed model outperforms the ADJSCC scheme with fewer FLOPs and a more significant number of parameters. Notably, within the SwinJSCC w/ SA\&RA variants model, SwinJSCC$^{B}$ w/ SA\&RA exhibits better performance compared to SwinJSCC$^{S}$ w/ SA\&RA and maintains considerable performance with SwinJSCC$^{L}$ w/ SA\&RA. Results indicate that the parameter amount of SwinJSCC$^{B}$ w/ SA\&RA has reached a saturation point, rendering the addition of parameters ineffective in performance improvement. Conversely, reducing the parameter count would lead to a significant decrease in performance.

	\section{Conclusion}\label{conclusion}
	
	In this paper, we have presented the establishment of a more expressive JSCC codec architecture that demonstrated the ability to adapt flexibly to diverse channel states and transmission rates within a single model. First, we have built an elaborate-designed neural JSCC codec based on the emerging Swin Transformer backbone, which achieves superior performance than conventional neural JSCC codecs built upon CNN while also requiring lower end-to-end processing latency. We have further upgraded our baseline SwinJSCC model to a versatile version by incorporating two design-specific spatial modulation modules. These modules scale latent representations based on the channel state information and target transmission rate, enhancing the model's capability to adapt to diverse channel conditions and rate configurations. Experimental results have shown that our SwinJSCC can achieve better or comparable performance versus the state-of-the-art engineered BPG + 5G LDPC coded transmission system with much faster end-to-end coding speed, especially for high-resolution images, in which case traditional CNN-based JSCC yet falls behind due to its limited model capacity.
	
	\ifCLASSOPTIONcaptionsoff
	\newpage
	\fi
	
	\bibliographystyle{IEEEtran}
	
	\bibliography{Ref}

\end{document}